\documentclass[aps,prd,showpacs,amsmath,amssymb,nofootinbib,eqsecnum, preprintnumbers, notitlepage,11pt]{revtex4-1}

\usepackage{geometry}
\usepackage{graphicx}
\usepackage[dvipsnames]{xcolor}
\usepackage{epstopdf}
\usepackage[abs]{overpic}
\usepackage{color}
\usepackage{comment}
\usepackage{soul}
\usepackage[colorlinks]{hyperref}

\newcommand{\be}{\begin{equation}}
\newcommand{\ee}{\end{equation}}
\newcommand{\ba}{\begin{eqnarray}}
\newcommand{\ea}{\end{eqnarray}}

\newcommand{\beq}{\begin{equation}}
\newcommand{\eeq}{\end{equation}}
\newcommand{\beqa}{\begin{eqnarray}}
\newcommand{\eeqa}{\end{eqnarray}}
\newcommand{\nn}{\nonumber}
\newcommand{\tcb}{\textcolor{blue}}

\newcommand{\rsm}{\rho_{sm}}

\newcommand{\Me}{{\cal{M}}}
\newcommand{\Ha}{{\cal{H}}}
\newcommand{\C}[1]{{\cal{#1}}}

\newcommand{\mean}[1]{\langle #1 \rangle}

\newcommand {\ket}[1]{\lvert \, #1\rangle}
\newcommand {\bra}[1]{\langle #1 \, \rvert}
\newcommand {\braket}[2]{\langle #1 \, | \, #2 \rangle}

\newcommand {\h}[1]{\hat#1{}}
\def \Tr{{\textrm{Tr}}}

\begin{document}
\title{Unitarity, Feedback, Interactions -- Dynamics Emergent from Repeated Measurements}

 \author{Natacha Altamirano}
\email{naltamirano@perimeterinstitute.ca}
 \affiliation{Perimeter Institute, 31 Caroline St. N. Waterloo
  Ontario, N2L 2Y5, Canada}
  \affiliation{Department of Physics and Astronomy, University of Waterloo,
  Waterloo, Ontario, Canada, N2L 3G1}

 \author{Paulina Corona-Ugalde}
 \email{pcoronau@uwaterloo.ca}
 \affiliation{Department of Physics and Astronomy, University of Waterloo,
 Waterloo, Ontario, Canada, N2L 3G1}
 \affiliation{Institute for Quantum Computing, University of Waterloo, Waterloo, Ontario, Canada, N2L 3G1}
 
 \author{Robert B. Mann}
 \email{rbmann@uwaterloo.ca}
 \affiliation{Department of Physics and Astronomy, University of Waterloo,
 Waterloo, Ontario, Canada, N2L 3G1}

 \author{Magdalena Zych}
 \email{m.zych@uq.edu.au}
 \affiliation{Centre for Engineered Quantum Systems, School of Mathematics and Physics, The University of Queensland, St Lucia, Queensland 4072, Australia}
 \date{\today}

\begin{abstract}
Motivated by the recent efforts to describe the gravitational interaction as a classical channel arising from continuous quantum measurements, we study what types of dynamics can emerge from a collisional model of repeated interactions between a system and a set of ancillae.  We show that contingent on the model parameters the resulting dynamics ranges from exact unitarity to arbitrarily fast decoherence (quantum Zeno effect). For a series of measurements the effective dynamics includes feedback-control, which for a composite system yields effective interactions between the subsystems. We quantify the amount of decoherence accompanying such induced interactions, generalizing the lower bound found for the gravitational example. However, by allowing multipartite measurements, we show that interactions can be induced with arbitrarily low decoherence. These results have implications for gravity-inspired decoherence models. Moreover, we show how the framework can include terms beyond the usual second-order approximation, which can spark new quantum control or simulation protocols.  Finally, within our simple approach we re-derive the quantum filtering equations for the different regimes of effective dynamics, which can facilitate new connections between different formulations of open systems.  
\end{abstract}

\pacs{03.65.Ta  03.65.Yz  04.60.-m}

\maketitle

\section{Introduction}
Modern measurement theory dispenses with the description of a measurement as a \textit{projection} onto one of the complete set of orthogonal eigensubspaces of a Hermitian operator (an observable) with the results (the observable's eigenvalues) distributed according to a probability measure \cite{VonNeumann:1932, Lueders:1951}. Rather, the measurement is understood as an \textit{operation}, whereby the system's final state is determined by an action of a \textit{completely positive trace non-increasing map}, corresponding to a given result, and the outcomes are described by linear operators on the system, distributed according to a \textit{positive-operator valued measure} (POVM) \cite{Kraus:1983}. 
This generalized description of a measurement allows achievement of tasks that are impossible with projective measurements \cite{NielsenBook2000} and is in fact necessary in most practical situations, where measurements are made with inefficient detectors, additional noise, or provide limited information about the system \cite{WisemanMilburn:Book:2010, jacobs2014BookQuantum}. 

Of key importance is that the POVM approach unifies the theory of measurements with a general description of dynamics, the theory of open quantum systems \cite{ref:BreuerBook}. It follows from Stinespring's dilation theorem \cite{Stinespring:1955} that any POVM operator can be constructed from a projective measurement on an enlarged Hilbert space: where the system of interest and an additional \textit{ancilla} evolve under a joint unitary and then the ancilla is measured. In the context of   measurement theory, the ancillae can be regarded as the measuring apparatus, whereas in the theory of open systems they can model the system's environment. Engineering a particular measurement and engineering a particular dynamics for the system are thus two complementary aspects of the same conceptual framework.
This correspondence is directly applied in quantum simulations~\cite{Lloyd1996}, quantum control  \cite{WisemanMilburn:Book:2010}, quantum computation \cite{Ladd2010QCNature, Du.2009decoupling} -- in all scenarios where a particular Hamiltonian for the system is desired, or when an existing system-environment interaction needs to be suppressed \cite{ViolaKnillLlyod:1999DD}. 

Recently, a particular model was developed where repeated position measurements result in an effective long-range \textit{interaction} between systems measured by common ancillae \cite{2013arXiv1311.4558K}. The interactions arise with dissipation of just the right magnitude to render the resulting dynamics classical --  unable to increase entanglement. 
The picture of interactions as mediated by quantum systems,  \cite{WeinbergQFT:1995}, is still missing for the gravitational case, despite a variety of efforts \cite{Kiefer:2014sfr}. The above result is thus of high interest \cite{Kafri:2014zsa, Kafri:2015iha, Tilloy:2015zya} for gravitational quantum physics. 
So far, an approximately Newtonian interaction was constructed from this model \cite{Kafri:2014zsa, Kafri:2015iha}, where decoherence does not only keep the resulting force classical, but is also claimed to be equivalent \cite{Tilloy:2015zya, Kafri:2014zsa, Wehner:2016UnivesalTest} to the Diosi-Penrose decoherence model \cite{Diosi:1986nu, ref:Diosi1989}.
However, it is also well known that any local dynamics can be efficiently simulated by suitably chosen interactions with ancillae \cite{NielsenBook2000}. In particular, repeated interactions employed in the research described above correspond  to a collisional model of an open system \cite{Rau1963, AlickiLendi1987Book, Ziman2005AllQubit, Ziman2005},  which can reproduce any Markovian dynamics \cite{Ziman2005AllQubit, Ziman2005} (including recently revisited examples of effectively unitarity \cite{Layden2015unitarity} as well as fully decoherent \cite{Layden2015QZE} evolutions). The questions thus arise: What are the assumptions necessary to obtain any particular type of dynamics from the continuous quantum measurement? Is it possible to induce the interactions but with less decoherence? Is it possible to generate an exact Newtonian, or even post-Newtonian, interaction from such a model? 
 
Here we study what types of dynamics can in general emerge from a simple model of repeated measurement, where refs.~\cite{2013arXiv1311.4558K, Kafri:2014zsa, Kafri:2015iha, Tilloy:2015zya} are a particular example. We show that the interaction terms found in those studies, arise for a  particular choice of the model parameters.  We discuss the necessary conditions and highlight all relevant assumptions required for their emergence. Furthermore, we quantify the amount of decoherence arising with the effective interactions. We provide a very simple proof that for bipartite measurements dissipation accompanying effective interactions is indeed lower bounded, generalizing the observation made in the gravitational example. However, we also show how effective interactions can emerge with arbitrarily low decoherence -- if one allows for measurements realized through, admittedly less appealing, multipartite system-ancillae interactions. 

While our results are motivated by  position measurements in the gravitational sector, they also have applicability beyond these particular considerations. 
The very simple approach applied throughout this work shows which assumptions can be modified, and how, in order to obtain a larger class of effective evolutions; for example, it  provides a means {to construct collisional models} that would give Markovian master equations beyond the usual Born-Markov approximation and  suggests how these can be used to recover exact Newtonian (or post-Newtonian) interaction terms from the repeated measurements.  
By deriving  quantum filtering equations corresponding to all the different regimes of emergent dynamics our work can also provide new connections between the stochastic calculus and other approaches to open quantum systems.
In this context we also note concurrent work \cite{DanDavid}
investigating emergent open dynamics of a quantum system undergoing rapid repeated unitary interactions with a sequence of ancillary systems.  Our results, are commensurate with these, though the work of ref. \cite{DanDavid} is concerned with understanding how thermalization, purification, and dephasing can emerge whereas  our concern is with the continuum limit and the nature of the emergent interactions arising in such models.

The structure of this paper is as follows: in Sec.~\ref{measurement} we revise a general model of a repeated  interaction between a system and a set of independent ancillae.  We show how  -- contingent on the relationship between the strength and duration of the  interaction and  the state of ancillae (moments of its probability distribution) --  any type of system dynamics can emerge:  from \textit{exact} unitary evolution (related to  ``decoherence free  subspaces'' \cite{Lidar2003}), effectively unitary evolution under an  ``external potential'' recently re-investigated in ref.~\cite{Layden2015unitarity} and decoherence, with  the quantum Zeno effect (QZE) \cite{Misra:1976by, 2002PhRvL..89h0401F, Layden2015QZE} in the extreme case. 
In Sec.~\ref{measure_feedback} we generalise the model to a \textit{sequence} of repeated
interactions.  In particular, we identify conditions under which coherent quantum feedback \cite{Lloyd2000Feedback, WisemanMilburn:Book:2010, jacobs2014BookQuantum} arises.
In Sec.~\ref{measurement_interaction}  we consider \textit{composite systems} under \textit{sequences} of  interactions. We identify conditions under which an effective interaction between two systems  emerges and quantify the accompanying  decoherence. For a particular choice of measurements we recover the emergence of the Newtonian gravitational interaction of ref.~\cite{Kafri:2014zsa}. Finally, we discuss the applied method, results and outlook in Sec.~\ref{discussion}, where we also discuss the connection to stochastic calculus.

 
\section{Continuous Quantum Measurement}
\label{measurement}

\begin{figure}[h]
\centering
\includegraphics[width=8cm]{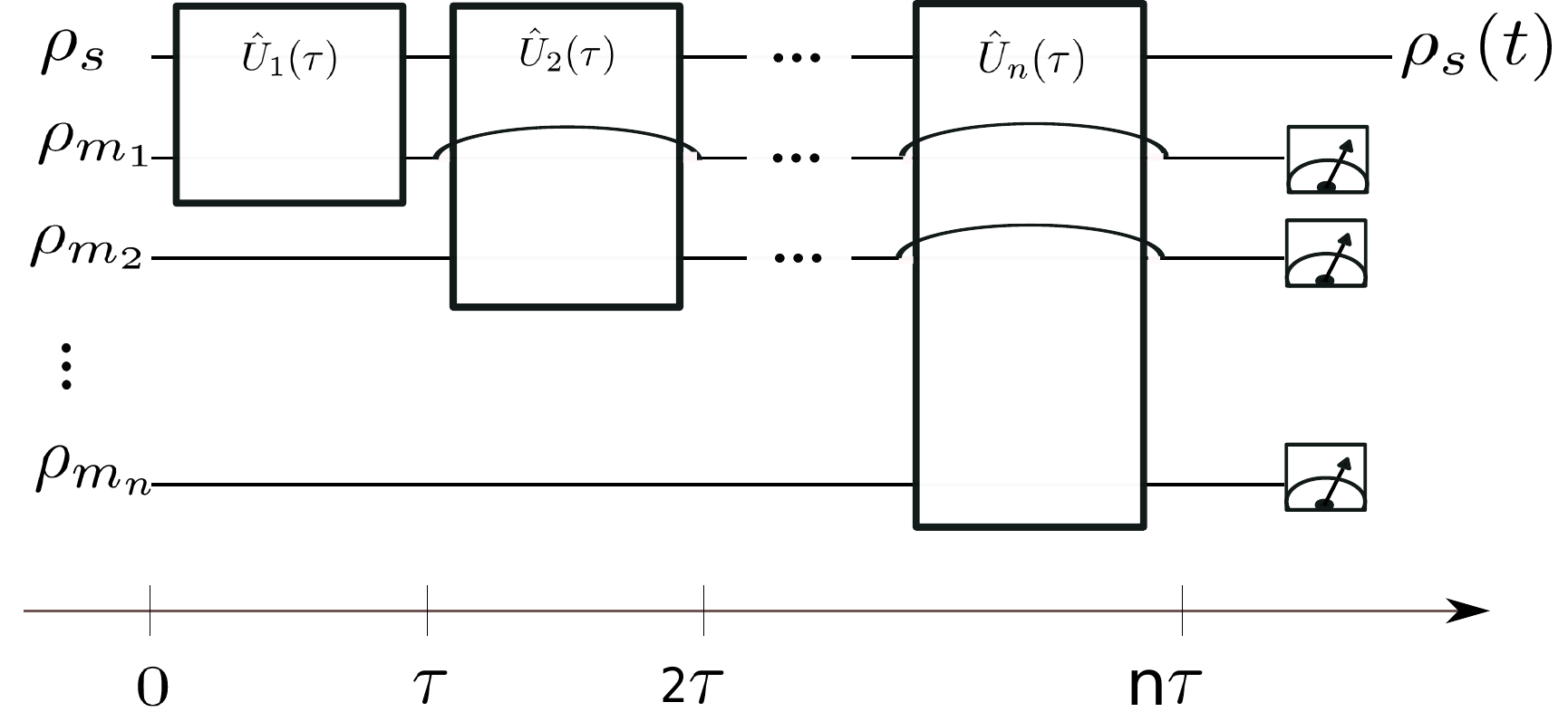}
\caption{Quantum circuit illustrating time evolution of a system subject to 
repeated interaction with $n$ ancillae. $\rho_s$ is the initial state of the system and $\rho_{m_i}$, $i=1,...,n$ -- of the $i^{th}$ ancilla. At each time step of duration $\tau$ the system interacts with one of the ancilla, the latter decouples and is discarded. Such a scenario is equivalent to a repeated measurement performed on the system by $n$ ``meters''. For identical $\rho_{m_i}$, the ancillae are also equivalent to a Markovian environment with relaxation time $\tau$. In the limit $\tau\to0$ the scenario describes continuous quantum interaction/measurement, or a memoryless (collisional) model of the system's environment.}
\label{circuit1}
\end{figure}

 
We consider a system $\C{S}$  and a set of $n$ identically prepared ancillae $\Me_r$, $r=1,...,n$. 
Initially, the system is uncorrelated with the ancillae, couples to the first one for a time $\tau$, decouples, then couples to the second one for time $\tau$, decouples, etc. This process  repeats $n$ times, as illustrated in Figure \ref{circuit1}. This is equivalent to a collisional model \cite{Rau1963, AlickiLendi1987Book, Ziman2005AllQubit, Ziman2005} of an open system, modelling interaction with a Markovian environment which has relaxation time $\tau$.
During an $r^{th}$ cycle the joint system $\C{S}\otimes\Me_r$ evolves under the Hamiltonian 
\beq
\h{\Ha}_{sm_r}=\h\Ha_0+g(t)\h\Ha_{I}=\h S_0 + \h M_0 +g_r(t) \h S\otimes \h M\,,
\label{Hamiltoniansm}
\eeq
where $\h S_0$ acts on the system only,  $\h M_0$  -- only on the ancilla and we thus call $ \C{\h H}_0:=\h S_0 + \h M_0 $ the total free Hamiltonian,  $\Ha_{I}:=\h S\otimes \h M$ is the interaction Hamiltonian. The latter is identical at each cycle: the same operators $\h S$ and $\h M$ act on $\C{S}$ and $\Me_r$ for each $r$ and the interaction strength satisfies $g_r(t)=g_{r+1}(t+\tau)$, where $\operatorname{supp}(g_r)=(t_r, t_{r+1})$ 
 and $t_{r+1}=t_r+\tau$. After the $r^{\text{th}}$ interaction the joint state of the system and the respective ancilla reads
\beq
\rho_{sm_r}(t_{r+1})=\h U_r(\tau)\rho_{sm_r}(t_r)\h U_r^\dagger(\tau)\,,
\label{evolutionsm}
\eeq 
where 
\beq
  \h U_r(\tau)={\cal{T}}\text{exp}\bigg(-\frac{i}{\hbar}\int_{t_r}^{t_r+\tau}{\cal{H}}_{sm_r}(t)dt\bigg)\,.
  \label{unitaryevolutionoperator}
  \eeq
For each interaction we assume the same initial state of the ancilla and as a result the final state of the system is described by $n$ iterations of a superoperator 
$\mathcal{V}(\tau)[\rho_s]:=\Tr_{\C{M}}\{\h U(\tau)(\h \rho_{s}\otimes \h \rho_{m}) \h U^\dagger(\tau)\}$, where $\Tr_{\C{M}}$ denotes the partial trace over the ancilla degrees of freedom,  and $\rho_{m}$ is the initial state of the ancilla. We are interested in the dynamics of $\rho_s$ in the limit of a continuous interaction, given by 
\beq
n\rightarrow\infty,\;\; \tau\rightarrow0, \;\;\;\; \mathrm{such\;that}\;\; \lim_{n\to\infty, \tau\to0}n\tau=T ,
\label{limit}\eeq
where $T$ is a fixed (and finite) time interval. Given the initial state of the system $\rho_s(0)$, the state at time $T$ is  fully described by  the map:
\beq
\rho_s(T)=\lim_{n\to\infty}\mathcal{V}^{n}\bigg(\frac{T}{n}\bigg)[\rho_s(0)]\,,
\label{general_map}\eeq
which is completely positive and trace preserving, but in general not unitary. The resulting dynamics in the continuous limit gives rise to a Markovian master equation, which we derive next.\\

If the interaction strength  $g_r(t)$ is continuous and differentiable in the interval $(t_r,t_r+\tau)$, the mean value theorem allows to obtain  
\be\label{mvt}
\int_{t_r}^{t_{r+1}}\C{\h H}_{sm}\,dt=(\C{\h H}_0+ \bar{g} \C{\h H}_{I})\tau,
\ee
 where $\bar{g}\!=\frac{1}{\tau}\int_{t_r}^{t_r+\tau}g_r(t)\,dt$ and the Hamiltonian
 $\C{\h H}:=\C{\h H}_0+\bar{g}\C{\h H}_{I}$ is  time independent. {Under certain restrictions on
the interaction strength $g_r(t)$, outlined in  Appendix \ref{app:higher_orders}, we can write the density matrix at a time $t_r$ as
\beq
 \h \rho_s(t_r)=\bigg( \h I+\sum_{m=1}^{\infty}{\cal{P}}_m\bigg)[\h \rho_s(t_{r-1})]\,,
  \label{operationondensity}
 \eeq
where ${\cal{P}}_m$ is the super-operator consisting of $m$ commutators:
 \beq
  {\cal{P}}_m[\h \rho_s(t_r)]=\frac{1}{m!}\bigg(\!\!-\frac{i \tau}{\hbar}\bigg)^m\mean{[\C{\h H},[\C{\h H},[...,[\C{\h H},\h \rho_s(t_{r-1})]]]]}_{\C{M}_r} ,
  \label{superoperator}
  \eeq
where $\mean{A}_{\C{M}_r}$ denotes the trace over the degrees of freedom of the r$^{\text{th}}$-ancilla.
Note that Eq.~\eqref{operationondensity} holds in particular for symmetric in time switching functions, e.g.~modelling interactions that are constant in time, and applies to typical scenarios involving photons, but also to toy models of gravitons in the recent gravitational decoherence models \cite{Kafri:2014zsa, Kafri:2015iha, Tilloy:2015zya} -- where we aim to apply results of this work.} 
Using Eq.\eqref{superoperator} to expand  \eqref{operationondensity}  yields
\ba
\rho_s(t_{n})&=&\rho_s(t_{n-1})-\frac{i}{\hbar}\tau[\h S_0+\bar{g}\mean{\h M}\h S,\rho_s(t_{n-1})] +\frac{i\tau^2}{2\hbar^2}\bar{g}\mean{i[\h M,\h M_0]}[\h S, \rho_s(t_{n-1})]+\nonumber \\
&-&\frac{\tau^2}{2\hbar^2}\bigg([\h S_0,[\h S_0,\rho_s(t_{n-1})]]
+\bar{g}\mean{\h M}[\h S,[\h S_0,\rho_s(t_{n-1})]]+\bar{g}\mean{\h M}[\h S_0,[\h S,\rho_s(t_{n-1})]]\bigg)+ \nonumber\\
&-& \frac{\tau^2}{2\hbar^2}\bar{g}^2\mean{\h M^2}[\h S,[\h S,\rho_s(t_{n-1})]]  +\cdots
\label{densityS}
\ea
where $\mean{\h M^k}\equiv\Tr_{\C{M}}\{\h M^k\rho_{m}\}$ for $k\in\mathbb{N}$.  Note that $i[\h M,\h M_0]$ is a Hermitian operator which can contribute to the effective unitary evolution of the system (see Sec.~\ref{sec:strong_int}). Analogous terms appear also at higher orders, and we discuss their potential contributions to the final master equation in Appendix \ref{app:higher_orders}. The equations of motion  for the system at time $T$ are finally obtained from 
 \be
 \dot{\rho}_s(T)=\lim_{\tau \to 0, n\to \infty}\frac{\rho_s(t_{n})-\rho_s(t_{n-1})}{\tau}.  
 \label{timederivative}
 \eeq
Equations \eqref{densityS} and \eqref{timederivative} define a general quantum master equation 
that describes the effect of repeated interactions with ancillae on the reduced state of the system. While  such collisional models are well studied in the context of open quantum systems and decoherence (see e.g.~refs.~\cite{Ziman2005AllQubit, Ziman2005}), the scope of the present work is to analyze the types of  \textit{unitary} contributions effectively arising in such models and to quantify their strength relative to the noise.

Equivalently, a repeated interaction of the form $\bar g\h S\otimes \h M$ describes a repeated 
measurement of the observable $\h S$ on the system made by the ancillae.  The
ancillae play the role of  ``meters'' (measuring apparatus) whose ``pointer states'' span a basis conjugate to the basis of the eigenstates of $\hat M$. The limit in Eq.~\eqref{limit} corresponds to a continuous measurement made over time $T$. Note, that since we work with a collisional model, we shall not consider  measurement channels that have no short-time expansion. %

The types of dynamics arising from such a continuous measurement in 
general depend on the interaction strength $g(t)$, the relation between the free and 
the interaction terms in the total Hamiltonian Eq.~\eqref{Hamiltoniansm} and on
the state of the ancillae. We shall discuss the different possibilities in the following section.

\subsection{Exact unitary evolution}
\label{exact_unitarity}

For an arbitrary initial state of the system the evolution under the Hamiltonian \eqref{Hamiltoniansm} is \textit{exactly} unitary if and only if:  $(i) $ the initial state of the ancilla is supported on a linear subspace $\mathcal{H}_M$ of eigenstates of $\h M$ with a common eigenvalue and $(ii)$ the subspace $\mathcal{H}_M$  is invariant under $\h M_0$. This is an analogous condition to the one derived in the context of decoherence free subspaces \cite{Lidar2003} or error correction \cite{PlenioVedralKnight.PRA.55.67}, with the crucial difference that here we present conditions on the state of the ancillae, rather than the system.  The proof is sketched in Appendix \ref{app:exact_unitarity}.

The conditions above, and the proof, naturally extend to the most general case of a bipartite interaction $\sum_{i=1}^{L}g_i\h S_i\otimes\h M_i$. The evolution of the system is exactly unitary if the joint state of the system and ancilla is supported on a subspace where  the total Hamiltonian can be written in block-diagonal form, where the system is in a joint eigenstate of a subset of operators $\h S_j$, with the corresponding eigenvalues $s_j$, and the ancilla is in an eigenstate of the operators $\h M_k$ in the remaining interaction terms, with eigenvalues $m_k$. The interaction then effectively reads $\sum_jg_js_j\h M_j + \sum_kg_k m_k \h S_k$. One also further requires that the free dynamics of the system and the ancilla preserve the above eigensubspaces. This generalizes the results discussed in \cite{Lidar2003} to an arbitrary interaction. The case of a general interaction, for a non-factorizable initial state, has also been studied in  \cite{Napoli.PRA.89.062104}.

From the viewpoint of the measurement interpretation of interactions such a scenario is somewhat unusual, since the allowed states of the system are constrained to a specific subspace (with the exception of a single interaction term, when only the state of the ancilla is constrained). The measurement interpretation can still be applied in the sense that time evolution of the measurement apparatus (ancillae) depends on the state of the system -- it is given by a Hamiltonian $\h h_m=\h M_0 +\sum_jg_js_j\h M_j$. Analogously, the system evolves under the Hamiltonian $\h h_s=\h S_0+\sum_kg_k m_k \h S_k$. The system's evolution is therefore on the one hand  ``interaction-free'' -- exactly unitary -- and on the other, it still depends on the state of another system, through the eigenvalues $m_k$.

\subsection{Effective unitarity}
\label{emergent_unitarity}

Unitary evolution of a system interacting with some environment typically emerges only as an approximate description -- when one assumes finite precision of any measurements made on the system to probe its dynamics. 
The quantitative condition for such an effective unitarity is that the terms $\propto \tau^k$ for $k\geq 2$ in Eq. \eqref{densityS} remain small compared to the first order ones, which is the case if
\beq\label{emergentu} 
 \lim_{\tau\to0}\frac{\tau^{k} \bar g^{k} \mean{\h M^k}}{\tau\bar g\mean{\h M}}=0\;,\;\;\; k=2, 3, \cdots
\eeq
These conditions are not automatically satisfied because the quantities $\mean{\h M^k}$  are moments of an in principle arbitrary probability distribution over the eigenstates of $\h M$ defined by the state of the ancilla\footnote{For finite-dimensional systems the number of independent moments is of course finite, as well as for continuous variable systems in e.g.~a Gaussian state, which yields distribution with only two independent moments.}.
If these conditions are met, from Eqs.~\eqref{densityS} and \eqref{timederivative} the following master equation is obtained\footnote{Terms containing at least one $\h M_0$ or $\h S_0$ in Eq.~\eqref{densityS} automatically have the required limiting behaviour: the expressions $\propto\!\tau^k$  with $\bar g^{k'}$ for $k'<k$ are small compared to lower order ones, $\propto\tau^{k'}\bar g^{k'}$, in the limit $\tau\to0$.}
\beq
\dot{\rho}_s=-\frac{i}{\hbar}[\h S_0+\Xi\h S,\rho_s],
\label{densityS2}
 \eeq
 where we defined 
 \beq
\Xi:= \lim_{\tau \to 0}\bar{g}\mean{\h M}.
 \eeq\label{Xi}
The system is effectively subject to an external potential  $\Xi\h S$ induced by the interactions and evolves approximately unitarily under an effective Hamiltonain $ \C{\h H}_{eff}=\h S_0+\Xi\h S$. The latter entails that in this regime the system-ancilla interaction is non-entangling.
Similar results have been found in the context of classical control theory of quantum systems \cite{Milburn2012PhilTransA}, where the system interacts with  ancillae that are themselves an open quantum system. The effective Hamiltonian $\C{\h H}_{eff}$ and the Hamiltonian in the case of the exact unitary dynamics $\h h_s$, Sec.~\ref{exact_unitarity}, have the same general structure but the key difference  is that $\C{\h H}_{eff}$  is valid only approximately, while $\h h$ holds exactly but only for a particular state of ancillae.

From the viewpoint of the measurement interpretation, the regime of effective unitarity is tantamount to a limiting case of a weak measurement  (or unsharp measurement~\cite{WisemanMilburn:Book:2010}) -- where the interaction between the system and the measuring apparatus is non-negligible only to lowest order.  Decoherence induced by such a measurement is vanishingly small, but so is the information about the system that could be gained from the apparatus, since each of the ancillae only evolves by a global phase -- as expected from general complementarity relations between information gain and state disturbance \cite{Busch2009}. 

An effective unitary evolution is a generic feature of a weak interaction regime: for $\lim_{\tau\to0} \tau\bar{g}=0 $ and a generic state of the ancilla -- with fixed but arbitrary moments $\mean{\h M^k}$ -- the reduced  state of the system evolves according to Eq.~\eqref{densityS2}. In this regime higher order corrections in $\tau\bar{g}$ can be made arbitrarily small by taking a suitably short time step $\tau$ (and therefore can be neglected provided that all subsequent measurements have finite resolution).

 Importantly, effective unitarity can also emerge in the strong interaction regime -- which we model by taking $\lim_{\tau\to0} \tau\bar{g}=C$ with $C=1$ for simplicity\footnote{For any value of $\tau$ a function $g_r$ satisfying  $\lim_{\tau\to0} \tau\bar{g}=1$ can be obtained from any suitably normalised  family of functions that converge to a Dirac delta distribution.}
 -- for specific states of the ancillae. The conditions in Eq.~\eqref{emergentu} now reduce to $\lim_{\tau\to0}\mean{\h M^k}/\mean{\h M}\to 0$, and we also need to ensure that $\Xi$ stays finite. An example of a suitable ancillae state is a Gaussian distribution over the eigenvalues of $\h M$ with mean $\alpha\tau$ and variance $\beta\tau$, where $\alpha, \beta$ are fixed parameters. The effective potential arising from this example is $\alpha\h S$.

\subsection{Quantum Zeno effect}
\label{sec:strong_int}

When at least one of the conditions in Eq.~\eqref{emergentu} is not satisfied, the reduced dynamics of the system is not unitary.  This can arise both in weak or strong interaction regimes, depending on the state of the ancillae.  We first focus on the regime of strong interactions, where non-unitarity will be shown to be a generic feature.

As in the section above, the strong interaction regime is understood as $\lim_{\tau\to0} \tau\bar{g}=1$.  We consider a generic state of the ancillae, where the moments $\mean{\h M^k}$ are in principle arbitrary but fixed, independent of $\tau$. 
The terms $\propto \tau^k\bar g ^k $ in Eq.~\eqref{densityS} will then  dominate over all others, and remain non-negligible in Eq.~\eqref{densityS} for arbitrary high $k$. Summing them all and denoting the magnitude of an arbitrary matrix element of the system (in the basis of $\h S$) by $\rho_{ij}:=|\bra{s_i}\rho\ket{s_j}|$, where $\h S\ket{s_i} = s_i\ket{s_i}$, yields
\ba\label{offdiagonals1}
{\dot\rho_{ij}}={\rho_{ij}}\lim_{\tau \to 0}\frac{1}{\tau}\left(|\mean{ e^{-i\frac{\Delta s_{ij}\h M}{\hbar}}}|-1\right),
\ea
with $\Delta s_{ij}:=s_i-s_j$. The right hand side remains finite only in two cases: for diagonal elements of the
system, $\rho_{ii}$, 
or for the ancilla in an exact eigenstate
of $\h M$.  For a generic state of the ancilla the suppression of the off-diagonal elements of the system becomes ``infinitely'' fast. More precisely,  $\mean{ e^{-i\frac{\Delta s_{ij}\h M}{\hbar}}}$
is a characteristic function of the probability distribution over the eigenvalues of $\h M$ defined by 
the state of the ancilla, and the moments of this distribution characterize the rate of decoherence. 
For a particular example of a Gaussian distribution $\mean{ e^{-i\frac{\Delta s\h M}{\hbar}}}=e^{-i\frac{\Delta s\mean{\h M}}{\hbar}}e^{-\frac{\Delta s^2\sigma^2}{2\hbar^2}}$ where we set $\Delta s\equiv\Delta s_{ij}$ for simplicity, and where $\sigma=\sqrt{\mean{\h M^2}-\mean{\h M}^2}$ is the variance of $\h M$, an exact solution to Eq.~\eqref{offdiagonals1} reads
\be\label{formsol}
{\rho_{ij}(t)}={\rho_{ij}(0)}\lim_{\tau \to 0}e^{{-\frac{t}{\tau}(1-e^{-{\sigma^2\Delta s^2}/{2\hbar^2}})}} ,
\ee
for $i\neq j$. Approximating \eqref{formsol} yields
\be\label{offdiagonals}
{\rho_{ij}(t)}\approx{\rho_{ij}(0)}\lim_{\tau \to 0}(1-\frac{ \sigma^2\Delta s^2}{2\hbar^2}\frac{t}{\tau}),
\ee
 to first non-vanishing order in $\sigma$. Eq.~\eqref{offdiagonals} is a generic result at this order -- valid for any state of ancilla when keeping up to second moments of its distribution.
 
In this regime, the interaction $\bar g\h S\otimes \h M$ is diagonalizing the system in the eigenbasis of $\h S$ at an arbitrarily fast rate 
 $\omega_D\approx\lim_{\tau\to0}\frac{\sigma^2\Delta s^2}{{2\hbar^2}{\tau}}$.   
Unitary evolution, stemming from the typically leading order term $\propto \mean{\h M}[\h S, \rho]$, becomes irrelevant -- it only acts non-trivially on the off-diagonal elements, but these are ``instantaneously'' suppressed. 
From the  viewpoint of the measurement-interpretation,  this ``infinite'' decoherence is simply the QZE effect: the measurements become repeated infinitely often  ($\tau\rightarrow0$) and projective (interaction strength diverges, $\bar{g}\propto1/\tau$) and the system ``freezes'' in the measurement basis.

Finally, note that the reduced dynamics of the system is non-unitary -- is discontinuous -- even for ancillae in an eigenstate of $\h M$ (with a non-zero eigenvalue), since $\dot\rho$ is then divergent. If only finite-precision measurements can be made on the system, decoherence and the QZE will arise also in that case.  Superposition states of the system will accumulate a relative phase at a divergent rate $\propto \lim_{\tau\to0}\mean{M}\Delta\!s/\tau$ and thus any coarse-graining will entirely suppress their coherence. Furthermore, assuming finite precision in the preparation of the ancillae (any non-vanishing variance)  decoherence will always be non-negligible in the strong interaction case -- and in this sense is a generic feature of the strong interaction regime. 

QZE has been realized with continuous (as well as pulsed) measurements e.g.~with Bose-Einstein condensates
\cite{PhysRevLett.97.260402}, and has been theoretically studied in a number of contexts, including freezing the evolution of a two-level Jaynes-Cummings atom interacting with a resonant cavity mode \cite{PhysRevA.45.5228},
controlling decoherence \cite{doi:10.1142/S0217979206034017}
producing effective hard-core repulsions in cold atomic gases \cite{Syassen1329,PhysRevLett.102.040402}, preparing and stabilizing the Pfaffian state in rotating harmonic traps loaded with cold bosonic atoms
\cite{PhysRevLett.104.096803}, and
inducing topological states of fermionic matter via suitably engineered dissipative dynamics
\cite{1367-2630-15-8-085001}.

\subsection{Finite decoherence}
\label{sec:finite_decoherence}

We now consider conditions under which only terms up to second order remain relevant. The model of repeated measurements reduces then to the usual Born-Markov master equation \cite{ref:BreuerBook}.  
In analogy to $\Xi$ defined in
Eq.~\eqref{Xi} we introduce
\be\label{meterlimits}
\qquad 
\Gamma:= \lim_{\tau \to 0}{\tau\bar{g}^2\mean{\h M^2}}\,,
\qquad  
\tilde M:=\lim_{ \tau\to 0}\frac{\bar{g}\mean{i[\h M,\h M_0]}}{2\hbar}\,,
\ee
 and assume that all higher order terms vanish in the considered limit (see also Appendix \ref{app:higher_orders}). This is indeed the case e.g.~(a) in a strong interaction regime ($\lim_{\tau\to0} \tau\bar{g}=1$) for ancillae in a Gaussian state with mean $\Xi \tau$ and variance $\sigma=\sqrt{\Gamma\tau-(\Xi \tau)^2}$; (b) in a \textit{weak} interaction regime (fixed $\bar g$) and ancillae in a Gaussian state with fixed $\mean{\h M}$ and $\mean{\h M^2}=\Gamma/\tau$. Importantly,  both in (a) and (b) the quantities $\Xi$, $\Gamma$ remain finite in the limit $\tau\rightarrow 0$. 
 
Eqs.~\eqref{densityS}, \eqref{timederivative}, \eqref{meterlimits} yield the following master equation
\be\label{master}
\dot{\rho}(t)=-\frac{i}{\hbar}[\h S_0+(\Xi-\tilde M)\h S,\rho]-\frac{\Gamma}{2\hbar^2}[\h S,[\h S,\rho]],
\ee
which features two different second order contributions: The term $\tilde M$ contributes to the unitary system dynamics and simply adds to the effective potential already present in Eq.~\eqref{densityS2}, and a  non-unitary term $-\frac{\Gamma}{2\hbar^2}[\h S,[\h S,\rho]]$, which results in decoherence at a finite rate. For the example (a) above, the off-diagonal elements of the system are suppressed according to 
\be\label{offdiagonals2}
\rho_{ij}(t)\approx\rho_{ij}(0)(1-\frac{ \Gamma\Delta\!s^2}{2\hbar^2}t), 
\ee
(neglecting $\h S_0$ for simplicity) in agreement with Eq.~\eqref{offdiagonals}.  
Decoherence vanishes provided that $\Gamma=0$ (implying $\Xi=0$) i.e. for an exact eigenstate of $\hat M$ with the eigenvalue 0, in agreement with the condition found in the QZE case (since the mean in the present example vanishes faster than the variance, unless $\mean{\h M^2}\equiv0$). 

The regime where Eq.~\eqref{master} applies and finite decoherence is observed corresponds to the typical case of  continous weak measurements: the interactions between the system and the measuring apparatus are finite but the contributions stemming from $\Gamma$ are considered non-negligible. In that context one often considers ancillae with trivial free evolution, $\tilde M=0$. The ensuing system dynamics features finite decoherence, due to noise introduced by the measurements, but with no modifications to the unitary part. 
 
As an exemplary application of the above, for a strong interaction and a  particular choice of operators: $\h M_0=0$, $\h M=\h{p}$ (momentum operator of the ancillae), $\h S=\h x$ (position operator of the system), and for a Gaussian state of the ancilla with $\mean{\h M}=0$ and $\mean{\h M^2}=\tau/D$, where $D$ is a fixed parameter, our Eq.~\eqref{master} reduces to a continuous position measurement derived in ref.~\cite{PhysRevA.36.5543}.

\section{Continuous measurement of multiple observables}
\label{measure_feedback}

\begin{figure}[h]
\centering
\includegraphics[width=10cm]{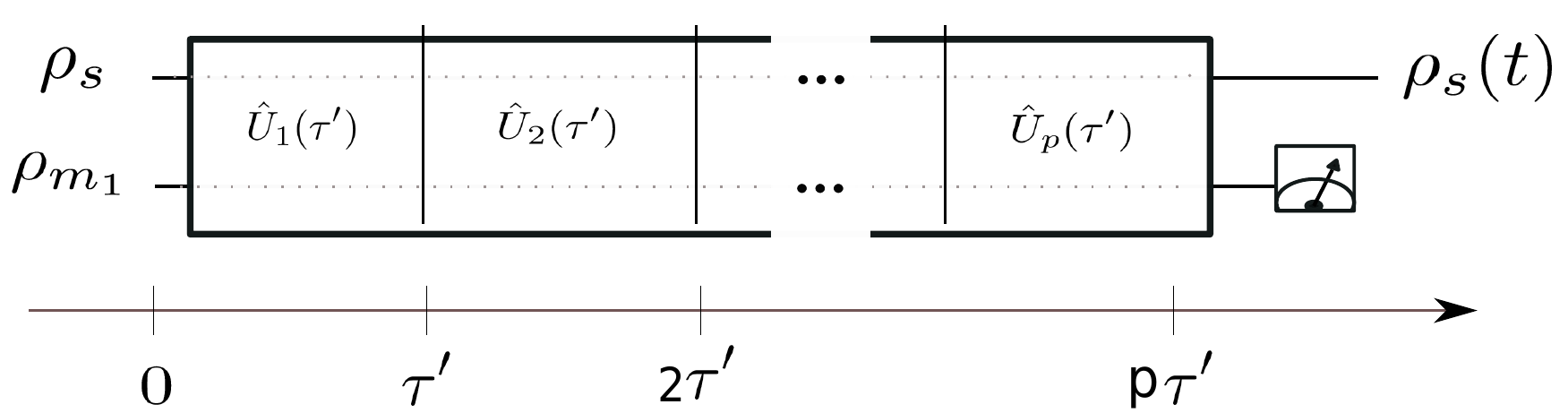}
\caption{Quantum circuit illustrating time evolution of a system $\C{S}$ during the first cycle of a repeated interaction with the ancillae. 
During each cycle, the system is subject to a sequence of $p$ different interactions with the same meter. All the subsequent cycles have the same structure.}
\label{fig2}
\end{figure}

Here we generalize our discussion to the case when several observables are repeatedly measured on the system. This situation can be accommodated by considering that each interaction in Sec.~\ref{measurement} is composed of $p$ sub-interactions, each of duration $\tau'=\tau/p$, as shown in Figure~\ref{fig2}. 
The total Hamiltonian in the $r^{th}$ cycle, Eq.~\eqref{Hamiltoniansm}, now generalizes to
 \beq\label{pcycleHamiltonian}
\C{\h H}^{(p)}_{sm_r}=\C{\h H}_0+\sum_{i=1}^pg_i(t)\C{\h H}^I_i=\h S_0 + \h M_0 +\sum_{i=1}^{p} g_i(t) \h S_i\otimes \h M_i.
\eeq
The operators $\h S_0, \h S_i$, $i=1,..., p$ act only on $\C{S}$, 
and $\h M_0, \h M_i$ act on the $r^{th}$ ancillae,  $g_i(t)$ is the switching 
function, now supported in the $i^{th}$ sub-step (of length $\tau'$), continuous in the interval where applied. The density matrix of the joint system at time $t_{r+1}$ is given by 
\beq
\rsm(t_{r+1})=\prod_{i=p}^{1}\h U_i(\tau')\rsm(t_r)\prod_{i=1}^{p} \h U_i(\tau')^\dagger\,,
\label{evolution-rsm-n}
\eeq
where $U_i(\tau')=e^{-\frac{i}{\hbar}\int_{t_r}^{t_r+\tau'}(\C{\h H}_0+g_i(t)\C{\h H}_i^I)dt}$.
 We apply the mean value theorem, as in Eq.~\eqref{mvt}, and define $\bar{g}_i=\frac{1}{\tau'}\int_0^{\tau'}g_i(t)dt$. Expanding Eq.~\eqref{evolution-rsm-n}  in powers of $\tau$ at time $t_n$  and tracing over the ancillae degrees of freedom gives
 \ba
\rho(t_{n})&=&\rho-  \sum_{i=1}^p\bigg\{\frac{i}{\hbar}\tau'[\h S_0+\bar{g}_i\mean{\h M_i}\h S_i,\rho] +\nonumber \\
&-&\frac{\tau'^2}{2\hbar^2}\sum_{j=1}^i(2-\delta_{ij})\bigg[[\h S_0,[\h S_0,\rho]]+\bar{g}_j\mean{\h M_j}
[\h S_0,[\h S_j,\rho]]+\bar{g}_i\mean{\h M_i}
[\h S_i,[\h S_0,\rho]] +\bar{g}_i\mean{[\h M_i,\h M_0]}
[\h S_i,\rho]\nonumber \\
&+&\frac{\bar{g}_i\bar{g}_j}{2}
\bigg(\mean{[\h M_i,\h M_j]}[\h S_i,\h S_j\rho + \rho \h S_j]+\mean{\{\h M_i,\h M_j\}}[\h S_i,[\h S_j,\rho]]\bigg)\bigg]\bigg\}\cdots\,,
\label{densitySn}
\ea
where 
$\{\h A,\h B\}:=\h A \h B +\h B \h A$ and $\rho\equiv\rho(t_{n-1})$.  
Equation \eqref{densitySn} generalises Eq.~\eqref{densityS} to the series of $p$ repeated measurements. It introduces a new type of term 
\beq
\mean{[\h M_i,\h M_j]}[\h S_i,\h S_j\rho + \rho \h S_j],
\label{feedback_term}
\eeq
which can contribute to the unitary part of the system dynamics. In particular, it can allow for feedback control of the system, discussed in Sec.~\ref{sec:feedback}. 

\subsection{Exact and effective unitarity}
\label{sec:pcycle_unitarity}

The conditions for exact unitary evolution of the system under arbitrary bipartite interactions with ancillae were discussed in Sec.~\ref{exact_unitarity} and they thus apply also to the present case, where the different interactions are applied sequentially.

The conditions for effective unitarity, Eq.~\eqref{emergentu}, directly generalize to the series of interactions.  The resulting effective dynamics reads
   \beq
 \dot{\rho}(t)=-\frac{i}{\hbar}[\h S_0 +\frac{1}{p}\sum_{i=1}^p\Xi_i\h S_i,\rho(t)],
 \eeq
where 
\beq\label{Xii}  
\Xi_i:=\lim_{\tau' \to 0}{\bar{g}_i\mean{\h M_i}}.
\eeq
This is a straightforward generalization of  Eq.~\eqref{densityS2}. The examples of interaction strengths and ancilla states discussed in Sec.~\ref{emergent_unitarity} apply to the present case as well. Thus, for multiple measurement/interactions effective unitary dynamics are also a generic feature of a weak interaction regime,  $\tau\bar g_i\to0$ for $i=1,..., p$,  but can also arise in the strong interaction regime for $\tau$-dependent state preparation of the ancillae.

 \subsection{Generalized QZE}
 \label{sec:pcycles_decoherence}

Here we consider the case when arbitrary high order terms contribute to the reduced dynamics of the system. Such situation arises in the regime of strong interactions, $\lim_{\tau\to0}\tau'\bar{g}_i=1$, $i=1,2$,  for a generic state of the ancillae. For clarity, below we restrict to $p=2$ sub interactions. 

In a full analogy to the QZE discussed in Sec.~\eqref{sec:strong_int}, the free evolution can be neglected compared to the interaction terms. Thus, the time evolution of the matrix  elements of the system reads
\ba\label{offdiagonalsp2}
\dot\rho_{ij}=\lim_{\tau \to 0}\frac{1}{\tau}\left(|\Tr_m\{\bra{s_i}e^{-i\frac{\tau'}{\hbar}\bar{g_2}\C{\h H}_2^I}e^{-i\frac{\tau'}{\hbar}\bar{g_1}\C{\h H}_1^I} \rho_m\otimes \rho e^{i\frac{\tau'}{\hbar}\bar{g_1}\C{\h H}_1^I}e^{i\frac{\tau'}{\hbar}\bar{g_2}\C{\h H}_2^I}\ket{s_j}\}|-\rho_{ij}\right).
\ea
%
%
As an illustrative example one can consider a repeated measurement of the same operator on the system $\h S_2=\h S_1\equiv \h S$ via two conjugate operators for the ancilla $[\h M_2, \h M_1]=i\hbar$.  Eq.~\eqref{offdiagonalsp2} then reduces to \be\label{offdiagonals1a}
{\dot\rho_{ij}}={\rho_{ij}}\lim_{\tau \to 0}\frac{1}{\tau}\left(|\mean{ e^{-i\frac{\Delta\!s_{ij}\h M'}{\hbar}}}|-1\right)
\ee
which is just \eqref{offdiagonals1} for $\h M':=\tau'\bar{g}_1\h M_1 +\tau'\bar{g}_2\h M_2$. Another  simple example is when conjugate observables are measured on the system (i.e.~$[\h S_2, \h S_1]=i\hbar$) via the same ancilla operator $\h M_1=\h M_2\equiv\h M$. Eq.~\eqref{offdiagonalsp2} then reads
\be\label{offdiagonals2}
{\dot\rho_{i'j'}}={\rho_{i'j'}}\lim_{\tau \to 0}\frac{1}{\tau}\left(|\mean{ e^{-i\frac{\Delta\!s'_{ij}\h M}{\hbar}}}|-1\right),
\ee 
where the off-diagonal elements are taken in the eigenbasis of $\h S':=\tau'\bar{g}_1\h S_1 +\tau'\bar{g}_2\h S_2 $,  defining  $\rho_{i'j'}:=\bra{s'_i}\rho\ket{s'_j}$ where $\h S'\ket{s'_i}=s'_i\ket{s'_i}$ and $\Delta s'_{ij}=s'_i- s'_j$.    
 In the most general case the decoherence basis is established from the full expression $$
e^{-i\frac{\tau'}{\hbar}\bar{g_2}\C{\h H}_2^I}e^{-i\frac{\tau'}{\hbar}\bar{g_1}\C{\h H}_1^I}=e^{-i\frac{\tau'}{\hbar}(\bar{g_1}\C{\h H}_1^I + \bar{g_2}\C{\h H}_2^I) -\frac{1}{2}\frac{\tau'^2}{\hbar^2} \bar{g_1}\bar{g_2}[\C{\h H}_2^I, \C{\h H}_1^I]+\cdots}\,.$$ 
Decoherence rates in this regime are again (cf.~Sec.~\ref{sec:strong_int}) formally divergent.

\subsection{Feedback}\label{sec:feedback}

Next we analyze conditions under which only terms up to second order contribute to the system dynamics. At the end of this section we discuss sufficient conditions for the emergence of feedback-control from the repeated measurement model.

To simplify the notation, along with $\Xi_i$, Eq.~\eqref{Xii}, we define
 \be\label{meterlimits_p2}
\qquad 
 \Gamma_{ij}:=\lim_{\tau' \to 0}\frac{1}{4}{\tau'\bar{g}_i\bar{g}_j\mean{\{\h M_i,\h M_j\}}}\,
 \qquad
\tilde M_{ij}:= \lim_{\tau' \to 0}\frac{1}{4\hbar}\tau'\bar{g}_i\bar{g}_j\mean{i[\h M_i,\h M_j]}\,.
\ee
From the Eqs.~\eqref{timederivative} and \eqref{densitySn} (with $\tau=2\tau'$) we obtain a general master equation for a system subject to two sequential measurements:
 \ba\label{master_p2feedback} 
 \dot \rho(t)&=&-\frac{i}{\hbar}[\h S_0+(\frac{1}{2}{\Xi_1}-\tilde M_{10})\h S_1+(\frac{1}{2}{\Xi_2}-3\tilde M_{20})\h S_2,\rho]+ \frac{i}{\hbar}\tilde M_{12}[\h S_2,\h S_1\rho + \rho \h S_1] +  \nonumber \\ 
 &-&\frac{1}{2\hbar^2} \sum_{i=1,2}\Gamma_{ii}[\h S_i,[\h S_i\rho]] -\frac{1}{\hbar^2}\Gamma_{12}[\h S_2,[\h S_1,\rho]]),  \ea
where in defining $\tilde M_{10}, \tilde M_{20}$ we introduced the convention $\bar{g}_0\equiv1$.

We now discuss how  coherent feedback  can result from the terms $\propto\tilde M_{12}$. Note that the first measurement $\h S_i\otimes\h M_i$ induces a translation of the state of the ancillae in the basis complementary to the eigenbasis of $\h M_i$. The  magnitude of this translation depends on the state of the system (on its $\h S_i$-eigenvalue).  The resulting state of the ancillae then determines the effective potential which arises for the system from the next interaction $\h S_j\otimes\h M_j$. Thus, for a suitable choice of the interactions and the state of the ancillae, an operation on the system is effectively 
performed that depends on its quantum state -- that is coherent feedback \cite{Lloyd2000Feedback, WisemanMilburn:Book:2010, Zhang2014quantum,  jacobs2014BookQuantum}. 
This makes clear why a necessary condition for feedback is $[\h M_i,\h M_j]\neq 0$. A sufficient condition is related with the question whether feedback is possible without introducing some decoherence. The answer is negative in the present model of ancillae.  The reason is that the feedback term $\frac{i}{\hbar}\tilde M_{12}[\h S_2,\h S_1\rho + \rho \h S_1]$ is \textit{at most} of the same order as the decoherence terms $ -\frac{1}{2\hbar^2}\sum_{i=1,2}\Gamma_{ii}[\h S_i,[\h S_i\rho]]$ -- a direct consequence of the inequality 
\be\label{feedback_noise_ineq}
\mean{(\bar g_1\h M_1-i\bar g_2 \h M_2)(\bar g_1\h M_1+i\bar g_2 \h M_2)}\geq 0.
\ee
Therefore, independently of the weak or strong interaction regime, the state of ancillae or the repetition rate of the measurements, with the present model of ancillae-system interactions, feedback-control of the system cannot be realized without introducing dissipation lower bounded according to Eq.~\eqref{feedback_noise_ineq}. See also refs.~\cite{Yamamoto:2014Feedback, Jacobs:2014feedback} for a comparison between coherent quantum feedback and the measurement-based feedback.

An example of a feedback-enabled control of a quantum system is a restoring force resulting from  a quadratic potential $\propto\hat S_1^2$. It can be achieved by taking $\h S_2 \propto  \h S_1$ in the model \eqref{master_p2feedback}.  More generally, feedback can take the form of a dissipative force, for $\h S_2 \propto \h S_1+\beta \h O$ for $ [\h O, \h S_1] \neq 0$. Taking canonically conjugate pair of ancillae operators $[\h M_i,\h M_j]\propto i\mathbb{I}$, results in feedback-control that is independent of the state of the ancillae.

Eq.~\eqref{master_p2feedback} is valid when the quantities $\Xi_i$, $\Gamma_{ij}$, $\tilde M_{ij}$ remain finite in the limit $\tau'\rightarrow0$, while contributions from higher moments vanish. A particular example of the ancillae state and operators that satisfy these conditions is a series of weak continuous position measurements first given in ref.~\cite{PhysRevA.36.5543}, see also Appendix \ref{app:examples}.  In this case, a harmonic potential arises as feedback and the accompanying decoherence keeps the momentum of the system finite. Experimental realization of feedback-control has been achieved with various systems, e.g. in cooling of optomechanical devices \cite{Poggio:2007cooling.cantilever}, trapped ions \cite{Bushev:2006coolions} or single atoms \cite{Koch:2010coolatom}.  

Finally, a tacit assumption was made in the above: that only measurements that are linear in the system operators can be realized by the ancillae. Relaxing this assumption would allow for noise-free feedback in the following sense:  If an arbitrary measurement/interaction was allowed -- of the form $\hat V\otimes\hat M$, for arbitrary  $\hat V$ -- one could induce an arbitrary potential term $\propto\mean{\hat M}\hat V$ already in the regime of \textit{effective unitarity}, Sec.~\ref{sec:pcycle_unitarity}.  For example, a quadratic potential arising due to weak measurement of the system position $\hat x$ in ref.~\cite{PhysRevA.36.5543} could be implemented unitarily if the ancillae would measure directly $\hat x^2$.  We note, however, that many experimental schemes (including optical devices \cite{BachorRalphBook2004QO}, mechanical oscillators \cite{BowenMilburnBook2015QOM}, atomic ensembles \cite{Hammerer:2010RevModPhys}) indeed allow only for such linear measurements/interactions.

\section{Measurement-induced dynamics for composite systems}
\label{measurement_interaction}

Here we consider the scenario from a previous section, but for a composite system. We allow that the different subsystems can have different interactions with the ancillae. We restrict our attention to a bipartite system subject to two continuously repeated interactions/measurements. 

For a system comprising subsystems $s_1, s_2$ the system operator describing the $i^{th}$ interaction in Eq.~\eqref{pcycleHamiltonian} most generally can be written as 
\be\label{compo_sys_operat}
\hat S_{i}=\sum_{j}c_j\hat S_{i,j}^{s_1}\otimes\hat S_{i,j}^{s_2}\;,
\ee
with real coefficients $c_j$, and where $\hat S_{i,j}^{s_{1(2)}}$ is an operator acting on subsystem $s_{1(2)}$.  As in the previous section, we are looking for a continuous limit of a protocol whose one step of duration $\tau$ is composed of two sub-steps, each of length $\tau'=\tau/2$. Thus, the master equation for such a case can directly be obtained from Eq.~\eqref{densitySn} for system operators given in Eq.~\eqref{compo_sys_operat}, and where the density matrix describes the state of both subsystems. 

The discussions in Sec.~\ref{measure_feedback} of the various regimes: unitarity (exact and approximate), QZE effect, finite decoherence, directly applies here. However the physical meaning of terms describing the induced potential, feedback, and decoherence is different: Since the operators $\hat S_i$ connect different subsystems, in general they entail emergence of interactions between them. Moreover, the decoherence basis will in general not be a product of the bases of the subsystems -- they can decohere into correlated states. This follows from the discussion of the decoherence basis in the QZE case of Eq.~\eqref{offdiagonals2} for system operators given by Eq.~\eqref{compo_sys_operat}.

Below we focus on a particular case where only bipartite interactions involving the ancillae are allowed -- i.e.~the ancillae only  interact with one subsystem at a time. This assumption has been made in the gravitational case studied in refs.~\cite{Kafri:2014zsa, Kafri:2015iha, Tilloy:2015zya} -- and is in fact crucial for the main results reported therein, as we will show at the end of this section.
Under the above assumption the system operators describing the interactions take the form: 
\be\label{product_operators}
\hat S_1=\hat S_1^{s_1}\otimes\C{\h I}^{s_2}\;,\;\;\; \hat S_2=\C{\h I}^{s_1}\otimes\hat S_2^{s_2}\;,
\ee 
where $\C{\h I}^{s_i}$ is the identity operator on the Hilbert space of subsystem $s_i$. Recall that  $\hat S_1$ acts in the first sub-step and $S_2$ in the second. For simplicity, below we take $\hat M_0=0$ (since $\h M_0\neq0$ would give terms analogous to those discussed in Sec.~\ref{sec:pcycles_decoherence}).
The total Hamiltonian acting during the entire $r^{th}$ interaction now reads
\begin{equation}\label{compositeHamiltonian}
\C{\h H}^{(p)}_{s_1s_2m_r}=\hat S_0+g_1(t)\h  {S_1}^{s_1}\otimes\C{\h I}^{s_2}\otimes \h {M_1} + g_2(t)\C{\h I}^{s_1}\otimes \h  {S_2}^{s_2} \otimes \h {M_2},
\end{equation}
analogously to the case of a single system in Eq.~\eqref{pcycleHamiltonian}. Operators  $\hat M_i$ act on the ancillae in the $i^{th}$ sub-step. 

\begin{figure}[h!]
 \includegraphics[width=8cm]{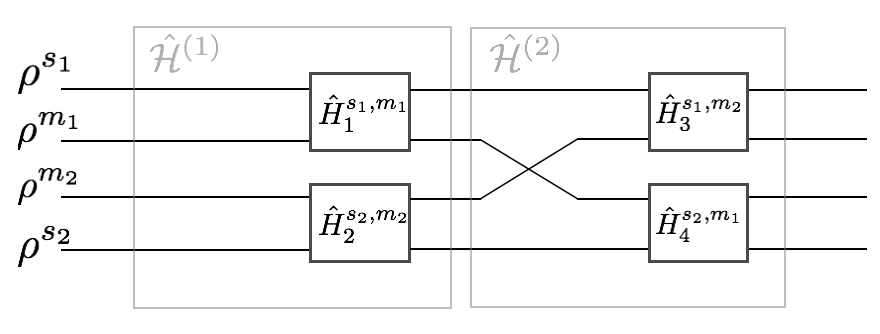}
 \caption{Composite system comprising subsystems $s_1, s_2$ prepared in the states $\rho^{s_1}, \rho^{s_2}$  interacting with ancillae $m_1, m_2$, initially in the states $\rho^{m_1}, \rho^{m_2}$. If each ancillae interacts with only one subsystem at a time, the resulting effective interaction between the subsystems is always accompanied by decoherence.}\label{circuitsubsystem}
\end{figure}

For the gravitational case it is natural to consider a symmetrized version of the above scenario: a second ancillae is added, which interacts with  $s_2$ in the first sub-step and with $s_1$ in the second sub-step. However, since this only doubles the terms already resulting from Eq.~\eqref{compositeHamiltonian} we defer the presentation of the symmetric case to the appendix \ref{app:example2}. In general, we can visualize the resulting process through the circuit in Figure \ref{circuitsubsystem}.  

From the Hamiltonian in Eq.~\eqref{compositeHamiltonian}, and with $\Xi_i, \Gamma_{ij}, \tilde M_{ij}$ defined in 
 Eqs.~\eqref{Xii},\eqref{meterlimits_p2} we obtain the master equation for a composite system under two consecutive continuous measurements:  
\ba\label{master_2systems}
 \dot \rho^{s_1,s_2}(T)&=&-\frac{i}{\hbar}[\h S_0+ \sum_{i=1,2}\frac{1}{2}{\Xi_i}\h S_i^{s_i},\rho^{s_1,s_2}]+ \frac{i}{\hbar}\tilde M_{12}[\h S_2^{s_2},\h S_1^{s_1}\rho^{s_1,s_2} + \rho^{s_1,s_2} \h S_1^{s_1}] +  \nonumber \\ 
 &-&\frac{1}{2\hbar^2} \sum_{i=1,2}\Gamma_{ii}[\h S_i^{s_1},[\h S_i^{s_i}\rho^{s_1,s_2}]] -\frac{1}{\hbar^2}\Gamma_{12}[\h S_2^{s_2},[\h S_1^{s_1},\rho^{s_1,s_2}]]),
 \ea
where ${\rho}^{s_1,s_2}$ is the joint state of $s_1, s_2$. Note, that this is a particular case of Eq.~\eqref{master_p2feedback} for the system operators defined in Eq.~\eqref{product_operators}.

We find in Eq.~\eqref{master_2systems} terms that are similar to those in Eq.~\eqref{master_p2feedback} for the single-system case. The terms $\propto\Xi_i$ describe effective potentials contributing to the unitary development that can arise with negligible decoherence, see Sec.\ref{sec:pcycle_unitarity}.  
The terms $\propto\Gamma_{ii}$ are decoherence terms for each subsystem. The term $\propto\Gamma_{12}$ in Eq.~\eqref{master_2systems}, analogous to the corresponding term in  Eq.~\eqref{master_p2feedback}, expresses the fact that the decoherence basis in general is given by some combination of the operators acting on the system in the different sub-steps. Note, however, that for to the operators in Eq.~\eqref{product_operators}, the resulting decoherence basis is still of a product form.  The term  $\propto\tilde M_{12}$, which more explicitly reads
\beq\label{effint}
\propto{\bar{g}_1\bar{g}_2\mean{[\h M_2,\h M_1]}}[\h S_2^{s_2},\h S_1^{s_1}\rho^{s_1,s_2}+\rho^{s_1,s_2}\h S_1^{s_1}],
\eeq
is analogous to the feedback term in Eq.~\eqref{feedback_term}.  However, the terms in Eq.~\eqref{effint} connect two subsystems and they thus introduce effective \textit{interactions} that can generate forces between them. The example of an approximately Newtonian interaction first derived in ref.~\cite{Kafri:2014zsa} is presented in Appendix \ref{app:example2}.

 For the  bipartite system-ancillae measurements, Eq.~\eqref{product_operators}, the effective interactions can only arise at the second  (or higher) order. As a result, the interactions terms are not larger than the decoherence terms arising from the double commutators, in full analogy to the case of feedback. The interactions are also lower-bounded by decoherence in the same way as feedback, as a consequence of the same inequality \eqref{feedback_noise_ineq}.  In fact, these effective interactions can also be interpreted as feedback:  the result of a measurement made by an ancillae on one system determines the strength of the effective potential acting on another system, which interacts with the same ancillae. 

However,  the conclusion about the necessary noise  does not arise if one allows more general measurements. A particular example is that of measurements realized simultaneously on both subsystems, described by system operators in Eq.~\eqref{compo_sys_operat}. In such a case the potential terms in Eq.~\eqref{master_2systems} would read  $\sim\Xi_i\sum_{j}c_j\hat S_{i,j}^{s_1}\otimes\hat S_{i,j}^{s_2}$ and could induce interactions even entangling the two systems. With such measurements the interaction terms could arise in the regime of \textit{effective unitarity}, Sec.~\ref{sec:pcycle_unitarity}, and would thus differ from the feedback schemes where the conditional state exhibits entanglement or where non-unitary terms are present \cite{WisemanMilburn:Book:2010, Zhang2014quantum, PhysRevA.71.042309, jacobs2014BookQuantum, PhysRevLett.98.190501}. The above assumption regarding interactions, and its role,  is fully analogous to the linearity assumption in the case of feedback  discussed at the end of Sec.~\ref{sec:feedback}. 
Finally, we note that the above scheme differs from  measurement-induced entanglement generation, where quantum correlations are created between systems interacting with a common environment by post-selecting on a particular state of the environment \cite{PhysRevLett.91.097905, chou2005measurement, PhysRevLett.112.170501, bernien2013heralded}. (Here, the environment is assumed to be inaccessible and is always averaged over.)


\section{Discussion}
 \label{discussion}
The very simple approach we have applied highlights several key aspects of the formalism. First, it stresses that a system subject to a repeated measurement/interaction with ancillae can still evolve unitarily or be subject to decoherence depending on the state of the ancillae, interaction strength and repetition rate (relaxation time of the environment). Second, in linear systems and under only bipartite system-ancillae interactions, the emergence of feedback-control and of induced interactions, respectively, is accompanied by a finite amount of decoherence, lower bounded by the magnitude of the induced unitary terms. However, decoherence  can be made arbitrarily low if  more general interactions with  the ancillae are permitted. For inducing interactions between different systems this would, however,  require non-local (multipartite) interactions.
 
Our approach has implications for generating gravitational interactions and gravitational decoherence. It has recently been shown that decoherence terms first introduced ad-hoc in gravity-inspired  models such as \cite{Diosi:1986nu}, could be derived from repeated interactions with the ancillae \textit{together} with an approximately Newtonian interaction \cite{Kafri:2014zsa, Tilloy:2015zya}. The intriguing aspect of this relation is that those two approaches are aimed at enforcing distinct notions of classicality. In decoherence models the desired classical regime is that where large spatial superpositions of massive systems are suppressed. Whereas in recent works, the notion of classicality is applied to interactions \cite{2013arXiv1311.4558K}, the classical regime being understood as eliminating the ability of interactions to generate entanglement. The approach of the present work can help clarify the extent to which these two notions of classicality have common consequences.

It would be quite remarkable if the \textit{exact} Newtonian or post-Newtonian interaction could be reproduced from the repeated-measurements model. The resulting theory could be seen as a toy-model for quantum gravitational degrees of freedom -- constructed not by quantizing their classical dynamics but by reconstructing the effective forces they generate. To this end, one would need to retain higher order terms than just first and second moments of the ancillae distribution (see appendix \ref{app:examples}). Our approach suggests a viable route in this direction:  one can ask whether a physical state of the ancillae exists that will have higher order non-vanishing moments such that the resulting unitary corrections to the system dynamics would sum to the Newtonian potential\footnote{For example, a skew Gaussian distributions have three independent moments. One can begin by asking what dynamics emerges if ancillae are prepared in such a skew-Gaussian state? Can one reconstruct a third-order approximation to the Newtonian potential within the measurement-based approach?}. As a further step one could extend the present approach beyond  Markovian processes  \cite{Pollock2015NM, Breuer:2016RMP} by incorporating, for example, initially correlated ancillae \cite{Rybar2012}, interactions between the ancillae  \cite{Ciccarello2013} or initially correlated system-ancillae states \cite{Modi:2012SciRep} -- particularly desirable for modelling gravitational degrees of freedom.  Finally, instead of constructing a classical channel from quantum degrees of freedom, one could ask if an entangling channel can arise from interactions with the ancillae in a scenario where the reduced state of the ancillae can nevertheless be described classically. The motivation here is that while the description of quantum states of matter in a general, even curved, space-time is well understood \cite{BIR82},  the problem lies in giving a consistent quantization of the latter. 

If it is indeed possible to generate general-relativistic gravity as an effective interaction with ancillae, the resulting toy-model of quantum-gravitational degrees of freedom could shed new light on the pernicious problems associated with quantum gravity. The above questions and in particular the practical question of detection of the interaction-induced decoherence and its implications for precision tests of gravity remain interesting subjects for further study.  

Departing from the main scope of the present work and allowing the ancillae to be accessible to an experimenter, we can describe the state of the system \textit{conditioned} on the results of measurements performed on the ancillae. The resulting  conditional dynamics takes the form of a quantum filtering equation.  Usually quantum filters  are studied using the quantum extension of the classical Ito calculus \cite{Bouten2007introduction,  Zhang2014quantum, WisemanMilburn:Book:2010, jacobs2014BookQuantum}. In Appendix \ref{app:QFiltering} we derive such equations from our simple model,  not only for the usual case of  finite decoherence (with feedback), but also for all other regimes discussed in the present work. We note that the usual quantum filtering approach assumes that measurement results are describable as a classical current and is thus less general than a fully quantum treatment, see e.g.~\cite{Jacobs:2014feedback, Yamamoto:2014Feedback}. In simple cases, where collisional models can be understood in terms of averaging the system dynamics over classical measurement outcomes, the two approaches are equivalent: averaging over all measurement records in the quantum filter formulation reduces the results to the unconditional evolution obtained in the main text. 

 \section{Conclusion}
 \label{conculsion}

Repeated interactions in the continuum limit are equivalent to the formalism of continuous weak measurements or collisional model of open systems.  This property can be exploited  to describe a broad range of phenomena, including  QZE, feedback control of a system,  emergence of  effective potentials or of an effective interaction between two systems subject to a measurement by the same apparatus/interacting with the same environment. Also ``interaction-free interactions" can emerge from such a model -- where evolution of a system is exactly unitary but where parameters of its Hamiltonian still depend on the quantum state of the ancillae with which it interacts.

The present approach provides a simple method for constructing collisional models of open systems beyond the Born-Markov approximation -- by considering ancillae states with higher order contributing moments (as suggested in the Discussion section for recovery of the exact Newtonian interaction). To this end, the model considered here requires taking a probability distribution with the desired number of independent moments. The resulting more general effective interactions can be beneficial in devising novel protocols for quantum control or error-correction and in new toy-models of gravitational degrees of freedom.

\section*{Acknowledgements}
The authors thank Fabio Costa, Gerard Milburn, Kavan Modi, Kiran Khosla and Felix Pollock for discussions. We thank the anonymous referee for motivating  inclusion of the perturbative  and quantum filtering analyses. This work was supported in part by the Natural Sciences and Engineering Research Council of Canada,  ARC Centre of Excellence for Engineered Quantum Systems grant number CE110001013 and the University of Queensland through UQ Fellowships grant number 2016000089. Research at Perimeter Institute is supported by the Government of Canada through Industry Canada,  by the Province of Ontario through the Ministry of Research and Innovation. P.C-U. gratefully acknowledges funding from CONACYT. N.A., P.C-U. and R.B.M.~are grateful for the hospitality of the University of Queensland where this work was initiated. 
M.Z.~acknowledges the traditional owners of the land on which the University of Queensland is situated, the Turrbal and Jagera people.


\appendix
\section{Magnus expansion and Higher order contributions}\label{app:higher_orders}

{In this section we discuss the condition on the terms in the total Hamiltonian and the switching function $g(t)$ under which equation \eqref{operationondensity} holds.}
\subsection{Magnus Expansion}

{Consider now Eq.\eqref{unitaryevolutionoperator}. Using time dependent perturbation theory we can write the evolution operator as a series expansion (Dyson series)
\beq
\h U_r(\tau)= 1+\sum_{k=1}^{\infty}P_k(\tau)\,, 
\label{Dyson_series}
\eeq
where
\beq
P_{k}(\tau)=\bigg(-\frac{i}{\hbar}\bigg)^k\int_0^{\tau}dt_1\,...\, \int_0^{t_{k-1}} dt_k\, {\cal{H}}_{sm_r}(t_1)\,...\,{\cal{H}}_{sm_r}(t_k)\,.
\eeq
From now on we denote $P_k$ to be of order $\delta t^k$. Equivalently, we can write the evolution operator in Eq.\eqref{unitaryevolutionoperator} using the Magnus Expansion \cite{CPA:CPA3160070404}
\beq
\h U_r(\tau)=\exp(\Omega(\tau))= 1+\sum_{k=1}^{\infty}\frac{1}{k!}\Omega^k(\tau)\,, 
\label{magnus_expansion}
\eeq}

{The terms in  \eqref{Dyson_series} and \eqref{magnus_expansion} can be related to each other if we write
\beq
\Omega=\sum_{m=1}^{\infty}\bigg(-\frac{i}{\hbar}\bigg)^m\Omega_{m}\,,
\eeq
where
\begin{eqnarray}
\Omega_1&=&\int_0^{\tau}{\cal{H}}_{sm_r}(t_1)dt_1\,, \nonumber \\
\Omega_2&=&\frac{1}{2}\int_0^{\tau}dt_1\int_0^{t_1}dt_2[{\cal{H}}_{sm_r}(t_1),{\cal{H}}_{sm_r}(t_2)]\,,  \\
\Omega_3&=&\frac{1}{6}\int_0^{\tau}dt_1\int_0^{t_1}dt_2\int_0^{t_2}dt_3([{\cal{H}}_{sm_r}(t_1),[{\cal{H}}_{sm_r}(t_2),{\cal{H}}_{sm_r}(t_3)]]+[{\cal{H}}_{sm_r}(t_3),[{\cal{H}}_{sm_r}(t_2),{\cal{H}}_{sm_r}(t_1)]]) \nonumber
\end{eqnarray}
Note, that $\Omega_k$ is also of order $\delta t^k$. Using this, one can show  that
\begin{eqnarray}
P_1&=&\Omega_1\,, \nonumber \\
P_2&=&\Omega_2+\frac{1}{2!}\Omega_1^2\,, \label{magnus_dyson} \\
P_3&=&\Omega_3+\frac{1}{2!}(\Omega_1\Omega_2+\Omega_2\Omega_1)+\frac{1}{3!}\Omega_1^3\,. \nonumber
\end{eqnarray}
Using Eqs.\eqref{Dyson_series} and \eqref{magnus_dyson} we can then write the evolution operator to order $\delta t^2$  as
\beq
\h U_r(\tau)=1+P_1+P_2=1-\frac{i}{\hbar}\Omega_1+\bigg(-\frac{i}{\hbar}\bigg)^2(\Omega_2+\frac{1}{2!}\Omega_1^2)=1-\frac{i}{\hbar}\Omega_1-\frac{1}{\hbar^2}\Omega_2-\frac{1}{2!\hbar^2}\Omega_1^2\,.
\eeq
If we now define 
\beq
\tilde{\Omega}_1=\frac{1}{\tau}\Omega_1\,,\,\,\,\,\,\,\,\,\,\,\tilde{\Omega}_2=\frac{1}{i\hbar\tau}\Omega_2\,,
\eeq
 We now can write the evolution operator as 
\beq
\h U_r(\tau)=1-\frac{i\tau}{\hbar}(\tilde{\Omega}_1+\tilde{\Omega}_2)-\frac{\tau^2}{2!\hbar^2}\tilde{\Omega}_1^{2}+{\cal{O}}(\tau^3)\,.
\label{evolution_magnus}
\eeq
Consider now the operator $\tilde{\Omega}_1$
\beq
\tilde{\Omega}_1=\frac{1}{\tau}\int_0^{\tau}{\cal{H}}_{sm_r}(t)dt=\C{\h H}_0+ \bar{g} \C{\h H}_{I}=\hat{\cal{H}}\,,
\eeq
where where have used Eq.\eqref{mvt} and $\hat{\cal{H}}$ is the Hamiltonian appearing in Eq. \eqref{superoperator}. It is staightforward to show  that Eqs.\eqref{operationondensity} and \eqref{superoperator} are equivalent to an expansion of equation \eqref{evolution_magnus} 
in powers of   $\tau$ that  neglects terms of order $\delta t^k$ relative to terms of order $(\delta t)^k$.
The net result is an expansion only in powers of  $\Omega_1^k\approx (\delta t)^k$. }

{Terms of order $\delta t^k$  will contribute if the Hamiltonian does not commute with itself at different times.  For example, at order $\delta t^2$ we find
\begin{eqnarray}
\tilde{\Omega}_2&\propto&\frac{1}{2}\int_0^{\tau}dt_1\int_0^{t_1}dt_2[{\cal{H}}_{sm_r}(t_1),{\cal{H}}_{sm_r}(t_2)]\,,  \\
&=&[\C{\h H}_0,\C{\h H}_I]\int_0^{\tau}dt_1\int_0^{t_1}dt_2[g(t_2)-g(t_1)]
\end{eqnarray}
which will vanish if either the free Hamiltonian commutes with the interaction part or if the integrals of the switching functions are zero (if $g(t)=g(\tau-t)$ in one cycle, at this order). }

{Finally, we note that by working in the interaction picture the Magnus expansion can be avoided because all time dependence is in the switching function $g(t)$, though it will then be necessary to transform back to the Heisenberg picture at the end of the calculation.  If the time dependence is in the Hamiltonian operator itself then the Dyson and Magnus expansions cannot be avoided, see e.g.~\cite{DanDavid} for a discussion of some of the resulting effects.}


\subsection{Higher order corrections}
In the present scenario the parameter $\bar{g}$ and the state of the ancillae  can depend on $\tau$. Thus in order to obtain the correct master equation for the system we need to examine which terms -- from any order of expansion of the evolution superoperator \eqref{superoperator} --   can give contribution of order $\tau$, as per Eq.~\eqref{timederivative}. We discuss this issue in detail below.

In general, order $\tau^k$ in the series \eqref{superoperator} 
\beq
\frac{1}{k!}\bigg(\!\!-\frac{i \tau}{\hbar}\bigg)^k\mean{[\C{\h H},[\C{\h H},[...,[\C{\h H},\h \rho_s]]]]}_{\C{M}}\,
\eeq
 contains terms with up to $k$ commutators of $\h S$ and $\h S_0$ with $\h \rho_s$; the number of $\h S$ operators in the commutator will give the power of $\bar{g}$ in the term.  As an example, we consider potentially relevant terms of order $k=3$:
\begin{eqnarray}
&&\bigg(\!\!-\frac{i \tau}{\hbar}\bigg)^3\bar{g}^3\mean{[\h H_I,[\h H_I,[\h H_I,\h \rho]]]}=\bigg(\!\!-\frac{i \tau}{\hbar}\bigg)^3\bar{g}^3\mean{\h M^3}[\h S,[\h S,[\h S,\h \rho_s]]]\,,\label{3com2}\\
&&\bigg(\!\!-\frac{i \tau}{\hbar}\bigg)^3\bar{g}^2\mean{[\h H_I,[\h H_I,[\h M_0,\h \rho]]]}=\bigg(\!\!-\frac{i \tau}{\hbar}\bigg)^3\bar{g}^2\mean{ [\h M^2,\h M_0]}[\h S,[\h S,\h \rho_s]]\,, \label{kk} \\
&&\bigg(\!\!-\frac{i \tau}{\hbar}\bigg)^3\bar{g}\mean{[\h S_0,[\h H_I,[\h M_0,\h \rho]]]}=\bigg(\!\!-\frac{i \tau}{\hbar}\bigg)^3\bar{g}\mean{[\h M,\h M_0]}[\h S_0,[\h S,\h \rho_s]]\,, \label{mtilde}\\
&&\bigg(\!\!-\frac{i \tau}{\hbar}\bigg)^3\bar{g}\mean{[\h H_I,[\h M_0,[\h M_0,\h \rho]]]}=\bigg(\!\!-\frac{i \tau}{\hbar}\bigg)^3\bar{g}\mean{[\h M_0,[\h M_0,\h M]]}[\h S,\h \rho_s]\,, \label{eff_uni}
\end{eqnarray}
(Note, that since free dynamics of the system is independent of $\tau$ no terms with only $\h S_0$, beyond the lowest order $k=1$, will contribute.)
The terms in Eqs.~(\ref{3com2}-\ref{eff_uni}) may be of order $\tau$ in specifically arranged scenarios. 
For example, consider the terms as in Eq.~\eqref{eff_uni} for an arbitrary order $k$ 
 \beq
\bar{g}\tau^{k}\mean{[\h M_0,[\h M_0,...[\h M_0,\h M]]]}[\h S,\h \rho_s]\,,
\label{M0_Mcommutators}\eeq
which has $(k-1)$ commutators between $\h M_0$ and $\h M$. 
These terms vanish in the strong coupling limit ($\lim_{\tau\to0} \tau\bar{g}=1$) but could survive for finite $\bar{g}$, for example, if $[\h M_0,\h M] = \lambda \h M$. (This may result in interesting dynamics, but we shall not consider such cases here.) Terms of the form \eqref{mtilde} will not contribute, which follows from Eq.~\eqref{meterlimits}. Finally, terms with $k$ commutators as in Eq.~\eqref{3com2} will be of order 
\beq
\tau^k \bar{g}^m\mean{\h M^m}\,,
\eeq 
where $0\leq m \leq k$ and $m$ is the number of $\h S$-operators in the commutator. If the condition in Eq.~\eqref{emergentu}  is met, the terms $\tau^k \bar{g}^k\mean{\h M^k}$ (factor with $k$ commutators of $\h S$ and $\h \rho$) vanish in the considered limit, but they do contribute in the strong interaction regime in the QZE case as discussed in Sec.~\ref{sec:strong_int}.   
Finally, let us note that in the case of ancillae implemented in a massive particle with $\h M_0\propto\h p^2$, which realises a  position measurements on the system, i.e.~$\h M=\h p$ and $\h S=\h x$,  expressions like Eq.\eqref{kk}-\eqref{eff_uni} and in particular all expressions in Eq.~\eqref{M0_Mcommutators} for $k\geq 2$ vanish. This is the case considered in Appendix \ref{app:examples}.

%

\section{Exact unitarity}\label{app:exact_unitarity}

Here we sketch the proof of the conditions that guarantee exact unitary evolution of a system whose dynamics is given in Eq.~\eqref{densityS}, discussed in Sec.~\ref{exact_unitarity}.  We repeat these conditions here for the convenience of the reader: The evolution of the system is \textit{exactly} unitary if and only if  $(i) $ the initial state of the ancillae is supported on a linear subspace $\mathcal{H}_M$ of eigenstates of $\h M$ with a common eigenvalue and $(ii)$ the subspace $\mathcal{H}_M$  is invariant under $\h M_0$.

It follows from $(i)$  that the total Hamiltonian ~\eqref{Hamiltoniansm}  acting on the joint state of the system and the ancilla can effectively be written as the sum of a term  $\h H_s(M)=\h S_0+\bar{g}M\h S$ that acts only on the system (where $M$ denotes the common eigenvalue of $\h M$ for states in $\mathcal{H}_M$),  and a term $\h M_0$ that acts only on the ancilla. From $(ii)$ we have that on the subspace $\mathcal{H}_M$ the operators $\h M_0, \h M$ commute. The evolution of the system thus factors out from the evolution of the ancilla, although it is described by an ancilla-dependent  Hamiltonian $\h H_s(M)$.

The proof of the necessary condition can be obtained as follows: A system evolves unitarily if its time evolution solves the Heisenberg equation 
\be
\dot{\rho}=-\frac{i}{\hbar}[\h h,\rho]
\ee for some Hermitian operator $\h h$. In particular, this means that to all orders in $\tau$  Eq.~\eqref{densityS} must agree order by order with a corresponding expansion of an equation of the form $\rho(t_n)=\h U_{h}(\tau)\rho(t_{n-1})\h U_h^\dagger(\tau)$ with $\h U_h(\tau)=\exp({-{i}\tau \h h}/{\hbar})$.  For this to be the case we must have $(a)$: $\h h=\h S_0+\bar{g}\mean{\h M}\h S$. Moreover, $(b)$ for all $k\in \mathbb{N}$ we must have $\mean{\h M^k} = \mean{\h M}^k$, which is equivalent to $(i)$ -- the state of the ancilla must be in an eigensubspace of $\h M$. Inspecting terms of order $\tau^3$, we obtain a further condition $(c)$: $\Tr\{(\h M\h M_0^2-2\h M_0\h M\h M_0 +\h M_0^2\h M)\rho_m\}=0$. Using $(i)$ and denoting by $M$ the corresponding eigenvalue of $\h M$, condition $(c)$ is equivalent to $M\Tr\{\h M_0\rho_m\h M_0\}=\Tr\{\h M\h M_0\rho_m\h M_0\}$, that is $\h M_0$ preserves $\mathcal{H}_M$. This completes the argument. \\

\section{Instantaneous position measurement and feedback-control}
\label{app:examples}

In this appendix we provide an example of Eq. \eqref{master_p2feedback} with specific operators satisfying 
the limits \eqref{meterlimits_p2}.
We take $\h S_0=\frac{\h p^2}{2m}$ as the free Hamiltonian of the system and  $\h M_0=0$ (trivial evolution of the meters). We take the first interaction to be effectively a measurement performed by the meter over the system of some operator $\h S_1$ for which we have $\h M_1=\h p_m$ where  $\h p_m$ is the momentum operator of the meter. For the second interaction we choose  $\h M_2=\h x_m$ where $\h x_m$ is the position operator of the meter. Finally, suppose we initially prepare each meter in a Gaussian state  $\ket{\psi}$ such that
\begin{equation}
\label{psi}
\psi(x) =\braket{\h x}{\psi}=\frac{1}{(\pi \sigma)^{1/4}} e^{-\frac{x^2}{2\sigma}}, 
\end{equation}
for which the density matrix is $\h \rho_m=\ket{\psi}\bra{\psi}$ and the various expectation values are
\begin{eqnarray}
\label{integrals}
 \mean{\h M_0}\!&=&\!0\,, \nonumber \\
 \mean{\h M_2}\!&=&\!\mean{\h x_m}=\!\int \!dx_m \bra{x_m} \h x_m \h \rho_m\ket{x_m}\!=\!\int\! dx_m x_m \psi(x_m)^2=0\, \nonumber \\
 \mean{\{\h M_1,\h M_1\}}\!&=&\!\mean{\{\h p_m,\h p_m\}}\!=\!2\!\!\int\!\! dx_m \bra{x_m} \h p_m^2 \h \rho_m\ket{x_m}\!=\!\!2\!\!\int\! \!dx_m \bigg[\bigg(\!\!i\hbar\frac{\partial}{\partial x_m}\!\bigg)^2\!\! \psi(x_m) \bigg]\psi(x_m)\!=\!\frac{\hbar^2}{\sigma}
\nonumber  \\
 \mean{\{\h M_2,\h M_2\}}\!&=&\!\mean{\{\h x_m,\h x_m\}}\!=\!2\!\!\int\! dx_m \bra{x_m} \h x_m^2 \h \rho_m\ket{x_m}\!\!=\!\int \!dx_m x_m^2 \psi(x_m)^2\!=\!\sigma \nonumber \\
  \mean{\{\h M_1,\h M_2\}}\!&=&\!\mean{\{\h x_m,\h p_m\}}=\int dx_m \bra{x_m} (\h x_m\h p_m+\h p_m \h x_m )\h \rho_m\ket{x_m}=0\, \nonumber \\
 \mean{i[\h M_1,\h M_2]}\!&=&\!\mean{i[\h x_m,\h p_m]}=-\hbar 
\end{eqnarray}

With these expressions the limits \eqref{meterlimits_p2} reduce to
\begin{equation}
  \Xi_i=0 \,\,\,\,\forall i\,, 
  \qquad 
 \Gamma_{11}=\lim_{\tau' \to 0}\frac{1}{4}{\tau'\bar{g}_1^2\frac{\hbar^2}{\sigma}}\,,
 \qquad 
 \Gamma_{22}=\lim_{\tau' \to 0}\frac{1}{4}{\tau'\bar{g}_2^2 \sigma}\,,
 \qquad 
 \tilde M_{12}=-\lim_{\tau' \to 0}\frac{\tau'\bar{g}_1\bar{g}_2}{4}\,.
\end{equation}

Let's now assume that the first interaction is an instantaneous measurement  that can be modelled by choosing its interaction strength $\bar{g}_1=\frac{1}{\tau'}$ -- which results in an interaction of the form of a delta function in time. The strength of the second interaction is constant over the interval $\tau'$ and finite.  The above limits  are thus \begin{equation}
  \Gamma_{11}=\frac{1}{4}\lim_{\tau' \to 0}\frac{\hbar^2}{\tau'\sigma}\,,
 \qquad 
 \Gamma_{22}=\frac{1}{4}\lim_{\tau' \to 0}{\bar{g}_2^2\,\tau' \sigma}\,,
 \qquad 
 \tilde M_{12}=-\frac{\bar{g}_2}{4}\,.
\end{equation}

We notice that in order to have a well defined master equation \eqref{master_p2feedback}
we need to take an initial state for the meter such that $\sigma$ goes to infinity in a way that $\lim_{\tau' \to 0}\sigma \tau'=D$, where $D$ is a constant. This is equivalent to considering a measurement which is instantaneous and ``infinitely'' strong but also ``infinitely'' inaccurate. Finally, the master equation \eqref{master_p2feedback} for our example reads
\begin{equation}
\label{master_feed}
\begin{aligned}
\dot{\rho}(t)={} &-\frac{i}{\hbar}[\h S_0, \rho(t)]-\frac{i}{\hbar}\frac{\bar{g_2}}{4}[\h S_2, \h  S_1\rho(t)+\rho(t)\h S_1]\\
& -\frac{1}{8D}[\h S_1, [\h S_1, \rho(t)]] -\frac{D}{\hbar^2}\frac{\bar{g}_2^2}{8}[\h S_2, [\h S_2,\rho(t)]]\,.
\end{aligned}
\end{equation}
Note that choosing $\h S_1=\sqrt{2}\h x_s$ to be the position operator of the system  and $\bar{g}_2=\sqrt{2}$ the last equation reduces to the unconditional  master equation 
\begin{equation}
\label{master_feed2}
\begin{aligned}
\dot{\rho}(t)={} &-\frac{i}{\hbar}[\h S_0, \rho(t)]-\frac{i}{2\hbar}[\h S_2,\h x_s\rho(t)+\rho(t)\h x_s]\\
& -\frac{1}{4D}[ \h x_s, [ \h x_s, \rho(t)]] -\frac{D}{4\hbar^2}[\h S_2, [\h S_2,\rho(t)]]\,.
\end{aligned}
\end{equation}
found by Milburn and Caves in ref.~\cite{PhysRevA.36.5543}. Those authors considered cycles of interaction, in which after an instantaneous position measurement a feedback control operation was introduced. 

In general, feedback-control gates are applied to a system to  control its behaviour 
after a measurement was performed. In the quantum information processing community, in the circuit framework, these are often approxiated as an instantaneous, non-infinitesimal transformations of the system depending on the measurement outcome. 
It is important to remark that contrary to this picture of feedback -- where the state of the meter is not  modified --  in the measurement/interaction approach applied here,  the state of the meter will naturally get modified. For the example above, it will have the net effect of shifting the momentum of the meter. Finally notice that this will not affect the results of short interactions in our regime since each ancillae is discarded after the measurement -- but it is an effect that needs to be taken into account when constructing measurement-based feedback gates.  For completeness, recall that if $\h S_2=\alpha \h S_1$, the second term in Eq. \eqref{master_feed} becomes a quadratic potential in $\hat S_1$.

\section{Emergent Newtonian-like interaction}
\label{app:example2}

Now we present a concrete application of Section \ref{measurement_interaction} by introducing the gravitational example of ref.~\cite{Kafri:2014zsa}.
We first genralize our model to the symmetric case, where the total Hamiltonians during the first and the second sub-step correspondingly read
\begin{eqnarray}
\C{\h H}^{(1)}&=&g_1(t)\h  {S_1}^{s_1}\otimes \h {M_1}^{m_1}\otimes \C{\h I}^{s_2} \otimes \C{\h I}^{m_2} +g_2(t)\C{\h I}^{s_1} \otimes \C{\h I}^{m_1}\otimes \h  {S_2}^{s_2}\otimes \h {M_2}^{m_2} 
\\
\C{\h H}^{(2)}&=&g_3(t)\h  {S_3}^{s_1}\otimes \C{\h I}^{m_1} \otimes \C{\h I}^{s_2} \otimes \h {M_3}^{m_2}+g_4(t)\C{\h I}^{s_1}\otimes \h {M_4}^{m_1} \otimes  \h  {S_4}^{s_2} \otimes\C{\h I}^{m_2} 
\end{eqnarray}

Following the same derivation as in the previous sections we obtain the master equation 
\ba
&\dot{\rho}&(T)= -\frac{i}{\hbar}[\h S_0, \rho] + \lim_{\tau'\to0}\nn \\
&-&\frac{i}{\hbar}\bigg(\bar{g}_2\mean{\h M_2^{m_2}}[\h S_2^{s_2},\rho]+\bar{g}_4\mean{\h M_4^{m_1}}[\h S_4^{s_2},\rho]+\bar{g}_1\mean{\h M_1^{m_1}}[\h S_1^{s_1},\rho]+\bar{g}_3\mean{\h M_3^{m_2}}[\h S_3^{s_1},\rho]\bigg)\nonumber\\
&-&\lim_{\tau\to 0}\frac{\tau}{4\hbar^2}\bigg(\bar{g}_2^2\mean{{\h M_2^{m_2}}^2}[\h S_2^{s_2},[\h S_2^{s_2},\rho]]+\bar{g}_4^2\mean{{\h M_4^{m_1}}^2}[\h S_4^{s_2},[\h S_4^{s_2},\rho]]\nonumber \\
&+&\bar{g}_1^2\mean{{\h M_1^{m_1}}^2}[\h S_1^{s_1},[\h S_1^{s_1},\rho]]+\bar{g}_3^2\mean{{\h M_3^{m_2}}^2}[\h S_3^{s_1},[\h S_3^{s_1},\rho]] \nonumber \\
&+&\bar{g}_2\bar{g}_3\mean{[\h M_3^{m_2},\h M_2^{m_2}]}[\h S_3^{s_1},\h S_2^{s_2}\rho+\rho\h S_2^{s_2}] + \bar{g}_1\bar{g}_4\mean{[\h M_4^{m_1},\h M_1^{m_1}]}[\h S_4^{s_2},\h S_1^{s_1}\rho+\rho\h S_1^{s_1}] \nonumber\\
&+&\bar{g}_2\bar{g}_3\mean{\{\h M_3^{m_2},\h M_2^{m_2}\}}[\h S_3^{s_1},[\h S_2^{s_2},\rho]]+\bar{g}_1\bar{g}_4\mean{\{\h M_4^{m_1},\h M_1^{m_1}\}}[\h S_4^{s_2},[\h S_1^{s_1},\rho]]\bigg)
\cdots\label{master_2systems}
\ea 
where ${\rho}$ denotes the state of both subsystems, $\h S_0$ is the free part of the Hamiltonian,  and we have suppressed the identity operators.

Since our main aim was to keep the terms in \eqref{effint}, which induce an interaction between the systems, we need the commutators $\langle[\hat{M_4}^{m_1},\hat{M_1}^{m_1}]\rangle$ and $\langle[\hat{M_3}^{m_2},\hat{M_2}^{m_2}]\rangle$ to be non vanishing. For simplicity, we choose:
\be 
\hat{M_1}^{m_1}=\hat{p}^{m_1}\quad
\hat{M_2}^{m_2}=\hat{p}^{m_2}\quad
\hat{M_3}^{m_2}=\hat{x}^{m_2}\quad
\hat{M_4}^{m_1}=\hat{x}^{m_1} 
\ee
and we use the ancillae initially prepared in a Gaussian state centered at zero as in Eq. \eqref{psi}. Using Eqns. \eqref{integrals} 
and assuming instantaneous interactions for $\bar{g}_2$ and $\bar{g}_1$ (such that $\bar{g}_i= \frac{\chi_i}{\tau}$) and constant interactions for $\bar{g}_3$ and $\bar{g}_4$ (such that $\bar{g}_3=\bar{g}_4=1$), then we get the following master equation:
\begin{equation}
\begin{aligned}
\dot{\rho}=&-\frac{i}{\hbar}[\h S_0, \rho] -\frac{i}{4\hbar}\left(\chi_2 [\hat{S}_3^{s_1},\hat{S}_2^{s_2}\rho+\rho\hat{S}_2^{s_2}]+ \chi_1 [\hat{S}_4^{s_2},\hat{S}_1^{s_1}\rho+\rho\hat{S}_1^{s_1}]\right)\\
&-\frac{D}{8\hbar^2}\left([\hat{S}_3^{s_1},[\hat{S}_3^{s_1},\rho]]+[\hat{S}_4^{s_2},[\hat{S}_4^{s_2},\rho]]\right)\\
&-\frac{1}{8D}\left([\hat{S}_1^{s_1},[\hat{S}_1^{s_1},\rho]]+\hat{S}_2^{s_2},[\hat{S}_2^{s_2},\rho]]\right)
\end{aligned}
\end{equation}

If we choose the operators $\hat{S}_i^{s_k}=\hat{x}_k^{s_k}$, and set $\chi_1=\chi_2 = K$, then this equation becomes
\begin{equation}
\begin{aligned}
\dot{\rho}=&-\frac{i}{\hbar}[\h S_0, \rho] -  \frac{i K}{4\hbar}\left([\h x_1,\h x_2 \rho+\rho\h x_2]+[\h x_2,\h x_1\rho+\rho\h x_1]\right)\\
&-\frac{D}{8\hbar^2}\left([\h x_1,[\h x_1,\rho]]+[\h x_2,[\h x_2,\rho]]\right) -\frac{1}{8D}\left([\h x_1,[\h x_1,\rho]]+[\h x_2,[\h x_2,\rho]]\right) \nonumber\\
=&-\frac{i}{\hbar}[\h S_0, \rho] -  \frac{i K}{2\hbar} [\h x_1\h x_2, \rho] -\frac{D}{8\hbar^2}\left([\h x_1,[\h x_1,\rho]]+[\h x_2,[\h x_2,\rho]]\right) -\frac{1}{8D}\left([\h x_1,[\h x_1,\rho]]+[\h x_2,[\h x_2,\rho]]\right) \nonumber\\
\end{aligned}
\end{equation}
which yields the quadratic potential for  an  induced gravitational interaction (with noise) studied in  \cite{Kafri:2014zsa}, taking  $\h S_0$ to be the sum of harmonic oscillator Hamiltonians for each subsystem.



\section{Quantum filtering equations}\label{app:QFiltering}
Below we derive quantum filtering equations from the collisional model, for different regimes of effective dynamics studied in the main text.

To obtain time evolution of a conditional state of the system,  instead of tracing over the ancillae we project it onto a state that corresponds to the measured outcome. Denoting the latter by $\ket{{x_n}}$ for the $n^{th}$ step,  instead of  \eqref{densityS} we get
\ba
\rho_c(t+\tau)&=&\mean{\rho_m}_n\rho_c(t)-\mean{\rho_m}_n\frac{i\tau}{\hbar}[\h S_0,\rho_c(t)] + \frac{\tau\bar{g}}{\hbar}\left(-i\mean{\hat M\rho_m}_n\h S\rho_c(t)+i\mean{\rho_m\hat M}_n\rho_c(t)\h S\right) \nonumber \\
&-&\frac{\tau^2\bar{g}^2}{2\hbar^2}\bigg(\mean{\h M^2\rho_m}_n\h S^2\rho_c(t) +  \mean{\rho_m\h M^2}_n\rho_c(t) \h S^2 -2\mean{\h M\rho_m\h M}_n \h S\rho_c(t)\h S\bigg)+ \cdots
\label{densityCondt}
\ea
where $\mean{\rho_m}_n\equiv \bra{{x_n}}\rho_m\ket{{x_n}}$ and where (following ref.~\cite{Zhang2014quantum}) we used the interaction picture for the ancillae.

The ancilla operator $\h M$ and its state determine the type of noise affecting the system.
Indeed, we can define a complex random variable with increment $d\mathcal{W}$ through 
\be\label{incrW}
-i\tau\bar{g}\mean{\h M\tilde \rho_m}_n=:\sqrt{\Gamma}d\mathcal{W},
\ee
 where $\tilde\rho_m:=\rho_m/\mean{\rho_m}_n$ (and the variable has distribution $\mean{\tilde\rho_m}_n$). Note, that $d\mathcal{W}$ is of dimension $\sqrt{T}$ and $\sqrt{\Gamma}$ is a dimensional constant.  The terms $\mean{\h M^2\tilde \rho_m}_n$ and $\mean{\tilde \rho_m\h M^2}_n$ in general differ from $\mean{\h M\tilde\rho_m\h M}_n$ (but the difference vanishes upon averaging over all measurement records). We take as the square of the stochastic increment the expression $\Gamma d\mathcal{W}d\mathcal{W}^*={\tau^2}\bar{g}^2 \mean{\h M\tilde\rho_m\h M}_n$, as this quantity is exactly equal to the square of the magnitude of the left-hand side of \eqref{incrW}  for pure $\rho_m$.

The real and the imaginary parts of $d\mathcal{W}$ can be expressed as 
\be\label{ReIm}
\Re\{d\mathcal{W}\}=\tau\bar{g}\mean{\frac{-i}{2}[\h M, \tilde \rho_m]}_n \qquad i\Im\{d\mathcal{W}\}=\tau\bar{g}\mean{\frac{i}{2}\{\h M,\tilde\rho_m\}}_n
\ee 
and so the probability for obtaining the outcome $x_n$, $\Tr{\rho_c}$, explicitly reads 
\be
\mean{\rho_m}_n\left(1+\frac{\sqrt{\Gamma}}{\hbar}\Re\{d\mathcal{W}\}\Tr{(\h S\rho_c+\rho_c\h S)}-\frac{\tau^2\bar{g}^2}{2\hbar^2}\mean{[\h M ,[\h M, \tilde\rho_m]]}_n\Tr\{S^2\rho_c\}\right)+...
\label{normalisation} 
\ee
Dividing Eq.~\eqref{densityCondt} by the probability in Eq.~\eqref{normalisation}, we obtain 
\ba\label{densityCondtNorm}
d\rho_c&=&-\frac{i}{\hbar}[{\tau\h S_0}+{\sqrt{\Gamma}} \Im\{d\mathcal{W}\}\h S,\rho_c] - {\frac{ \Gamma}{2\hbar^2}}[\h S,[\h S,\rho_c]]{d\mathcal{W}d\mathcal{W}^*} \\\nonumber
&+&\frac{\sqrt{\Gamma}}{\hbar}\Re\{d\mathcal{W}\}\left(\h S\rho_c+\rho_c\h S-\rho_c\Tr\{\h S\rho_c+\rho_c\h S\}\right)\\\nonumber
&-&\frac{\tau^2\bar{g}^2}{2\hbar^2}\bigg(\mean{\h M^2\tilde\rho_m -\h M\tilde\rho_m\h M}_n\h S^2\rho_c+\mean{\tilde\rho_m\h M^2 -\h M\tilde\rho_m\h M}_n\rho_c\h S^2\\\nonumber
&-&\rho_c\mean{[\h M ,[\h M, \tilde\rho_m]]}_n\Tr\{S^2\rho_c\}\bigg)+...
\ea
which describes the evolution of the normalized conditional density operator of the system. 

Upon averaging over the measurement outcomes  Eq.~\eqref{densityCondtNorm} becomes identical to Eq.~\eqref{densityS}. In particular the last two lines of Eq.~\eqref{densityCondtNorm}, which contain corrections stemming from the difference between the terms $\mean{\h M^2\tilde \rho_m}_n$, $\mean{\tilde \rho_m\h M^2}_n$ and $\mean{\h M\tilde\rho_m\h M}_n$,
vanish. More generally, these terms are equal provided the state of the ancillae is either diagonal in the eigenbasis  of $\h M$ or if $\ket{x_n}$ is an eigenstate of $\h M$. For simplicity we  ignore these corrections hereafter.

Equation \eqref{densityCondtNorm} is a quantum filtering equation \cite{Bouten2007introduction, Zhang2014quantum, WisemanMilburn:Book:2010, jacobs2014BookQuantum}, with noise given by a general process $d\mathcal{W}$. The real part $\Re\{d\mathcal{W}\}$ has vanishing average and gives the usual stochastic term, while the average of $\Im\{d\mathcal{W}\}$ is proportional to $\mean{M}$ and contributes to the unitary dynamics of the system.

In the Heisenberg picture Eq.~\eqref{densityCondtNorm} also gives the evolution of any system observable (up to a sign of $i$). Thus,  outcomes of measurements of  $S$ conditioned on the observations made on the ancilla exhibit noise given by $d\mathcal{W}$, up to order $\tau$: 
$${\dot S}= \mean{{\dot S}}_c^{free}+\frac{2\sqrt{\Gamma}\Delta S_c^2}{\hbar}\frac{\Re\{d\mathcal{W}\}}{dt}$$ where $\Delta S_c^2 =\mean{\h S^2}_c-\mean{\h S}_c^2$ is the conditional variance of $\h S$, see also \cite{Zhang2014quantum}. 

\subsection{Input-output relation for the ancillae}

Up to order $\tau$, an ancilla operator $\h m$ evolves as
\ba
\h m(t+\tau)=\h m + \frac{i\tau}{\hbar}[\h M_0, \h m]+ \frac{i\tau}{\hbar}[\h M, \h m]\h S
\label{anc_operator}
\ea
where for clarity we absorbed the factor $\bar{g}$ into $\h M$. In the usual case, when the measurement is done in a complementary basis to $\h M$,  we have 
\ba
\h m(t+\tau)=\h m(t) + {\tau}f\frac{i}{\hbar}\h M +\tau\h S,
\label{anc_operator}
\ea
provided  $[\h M_0, \h m]=f\h M$, for some dimensional (and possibly complex) $f$. (This holds e.g.~for position measurements  $\h m=\h x$ on the ancillae, which are assumed to be free particles with $M_0\propto \h p^2/2$ and which interact with the system via $\h M= \h p$, a relevant case for  our work).
Taking the conditional expectation values and defining the measurement current $\displaystyle{\dot m_c:=\lim_{\tau\to0}\frac{\mean{\h m(t+\tau)-\h m(t)}_n}{\tau}}$ we see that $\dot m_c$ provides a noisy measurement of $\h S$, from Eq.~\eqref{anc_operator}: 
\ba
 \mean{\h S}_c=\dot m_c + \frac{f\sqrt{\Gamma}}{\hbar}\xi, 
\label{meas_current} 
\ea
{where $\mean{\h S}_c$ is the conditional mean and $\displaystyle{\xi:=\lim_{\tau\to0}\frac{d\mathcal{W}}{\tau}}$.}


\subsection{Regimes of dynamics}
As stressed in the main text, depending on the interplay between the state of the ancillae, the interaction strength $\bar{g}$, and operator $\h M$, different types of dynamics emerge. This is also the case for the conditional dynamics  described by \eqref{densityCondtNorm}. The basis in which the ancillae are measured defines a so-called unravelling of the master equation (but give the same unconditional dynamics). 

\paragraph{Exact unitarity (Sec.~\ref{exact_unitarity}).} 
 The ancillae are here in an eigenstate of $\h M$, with a common eigenvalue $M$.   The conditional state of the system is given by
\ba
\rho_c(t+\tau)=\langle e^{-\frac{i}{\hbar} \bar{g} \h S\h M \tau}\rho(t)e^{\frac{i}{\hbar} \bar{g} \h S\h M \tau}\rangle_n
\ea
and the conditional master equation \eqref{densityCondtNorm} thus reads
\ba
\dot{\rho_c}=-\frac{i}{\hbar}[\h S_0+\Im\{\mathcal{\xi}\} \h S,\rho], 
\ea
with $\xi$ defined as in the section above. Since ${\langle \h M\tilde\rho_m\rangle_n}\equiv M$, the stochastic increment describes a process with exactly vanishing noise.

\paragraph{Effective unitarity (Sec.~\ref{emergent_unitarity}).}
{From eq.~\eqref{emergentu} directly follows that the noise increments satisfy $d\mathcal{W}= \mathcal{O}(\tau)$ and $d\mathcal{W}^2=\mathcal{O}(\tau^2)$.} Hence, only the first order terms contribute, and the stochastic master equation reads
\be
\dot\rho_c=  \lim_{\tau\to0}\frac{d\rho_c}{\tau}  =
- \frac{i}{\hbar}[\h S_0+\Im\{\xi\}\h S,\rho_c] 
+\frac{\sqrt{\Gamma}}{\hbar}\Re\{\xi\}\bigg(\h S\rho_c+\rho_c\h S-\rho_c\Tr\{\h S\rho_c+\rho_c\h S\}\bigg),
\label{stochU}
\ee
{In the collisional-model picture in the strong interaction regime ($\bar g\propto1/\tau$)  the ancilla must be close to an eigenstate of $\h M$, but for the weak interaction regime ($\bar g$ constant) the ancilla can be in a generic state, see discussion in the main text.
Both these cases consistently define stochastic variable with small variance, which results in a process with  \textit{effectively} vanishing noise. Note, that the last term describes fluctuations in the evolution of the system due to statistical fluctuations in the measurement outcomes on which the system dynamics is conditioned: Indeed, $\Re\{\xi\}\propto\mean{\frac{-i}{2}[\h M, \tilde \rho_m]}_n$ vanishes if $\rho_m$ commutes with $\h M$, or if the unravelling (measurement basis $\ket{x_n}$) is the eigenbasis of $\h M$.}

\paragraph{Finite decoherence (Sec.~\ref{sec:finite_decoherence}).}
The first of the conditions in Eq.~\eqref{meterlimits} implies that $d\mathcal{W}^2=\mathcal{O}(\tau)$, while the higher order terms vanish.  Thus the random variable corresponds to the Wiener process. The stochastic master equation takes the form
\ba
\dot\rho_c=
&-&\frac{i}{\hbar}[\h S_0+\Im\{\xi\}\h S,\rho_c]-\frac{\Gamma}{2\hbar^2}[\h S,[\h S,\rho_c]]\\ \nonumber
&+&\frac{\sqrt{\Gamma}}{\hbar}\Re\{\xi\}\bigg(\h S\rho_c+\rho_c\h S-\rho_c\Tr\{\h S\rho_c+\rho_c\h S\}\bigg).
\label{stochFiniteDec}
\ea
We note that ancilla e.g.~in a Gaussian state in a basis complementary to $\h M$, and with zero mean, defines an ideal Wiener process, with vanishing average of $\Re\{d\mathcal{W}\}$, and $\Im\{d\mathcal{W}\}$ identically zero. 

\paragraph{Zeno effect (Sec.~\ref{sec:strong_int}).}
In this case we have $\tau\bar{g}\rightarrow 1$ with fixed (and finite) moments $\mean{\h M^k}_n$. All powers of the increment $d\mathcal{W}$ thus contribute $\mathcal{O}(1/\tau)$ to the master equation. In a full analogy to Eq.~\eqref{offdiagonals1}, the leading terms give
\ba
\dot\rho_c=\lim_{\tau\to0}\frac{1}{\tau}\left(\sum_{n=0}^{\infty}
\left(\frac{-i}{\hbar}\right)^n d\mathcal{W}^n[\h S,[\h S ...[\h S, \rho_c]...]]-\rho_c\right),
\label{stochZeno}
\ea
where for simplicity we assumed a real stochastic increment.

\subsection{Feedback}
If the measurement outcomes are known, as the filtering approach assumes, after each measurement one can perform  feedback control on the system by applying a Hamiltonian based on the measurement result. The conditional state of the system,  discussed above, is then after each measurement additionally modified according to

\ba
\rho(t+\tau)=e^{-\frac{i}{\hbar}H_{FB}\tau}\rho_c(t)e^{\frac{i}{\hbar}H_{FB}\tau},
\label{feedbackStoch1}
\ea
with $H_{FB}=\bar{g}_2\dot{\chi} \h S_2$ with $\dot{\chi}$ the measurement current~\cite{Wiseman:1994feedback}.  Suppose\footnote{The dimensional factor ${\hbar}/\sqrt{\Gamma}$ is necessary for consistency of units, and an additional factor of $2$ is chosen to simplify the form of the final master equation. In general, the measurement current can be redefined up to a factor depending on the measurement procedure \cite{WisemanMilburn:Book:2010}.}
$\dot{\chi}=\langle\h S\rangle_c+\frac{\hbar}{2\sqrt{\Gamma}}\frac{d\mathcal{W}}{dt}$, see also Eq.~\eqref{meas_current}, and we restrict for simplicity to real $d\mathcal{W}$.
For the case corresponding to finite decoherence, we take in Eq.~\eqref{feedbackStoch1} the conditional state given by Eq.~\eqref{stochFiniteDec} and expand the feedback operators ignoring second order terms and higher, which will vanish in the limit $\tau\to 0$. The resulting master equation for a conditional dynamics including the continuous measurement and feedback process is 
\ba
\dot{\rho}_c=&-\frac{i}{\hbar}[\h S_0,\rho]-\frac{\Gamma}{2\hbar^2}[\h S,[\h S,\rho]]-\frac{\bar{g}_2^2 }{8 \Gamma}[\h S_2,[\h S_2,\rho]]+\frac{-i\bar{g}_2 }{2\hbar}[\h S_2,\h S\rho+\rho\h S] \nonumber \\
&+\frac{d\mathcal{W}}{dt}\left(\frac{\sqrt{\Gamma}}{\hbar}(\h S\rho+\rho\h S-2\langle\h S\rangle\rho)-\frac{i\bar{g}_2}{2\sqrt{\Gamma}}[\h S_2,\rho]\right).
\label{Stochastic_feedback}
\ea
Averaging over the measurement results, only the first line of Eq.~\eqref{Stochastic_feedback} remains. The resulting equation has the same form as the collisional model feedback in Eq.~\eqref{master_p2feedback}.

\bibliographystyle{linksen}
\bibliography{Databaze}
\end{document}